\documentclass[12pt]{article}
\begin{document}

\title{Asymptotic correlation functions of \\ Coulomb gases on an annulus}

\author{Taro Nagao} 

\date{} 

\maketitle

\begin{center} 
\it Graduate School of Mathematics, Nagoya University, Chikusa-ku,
\\  Nagoya 464-8602, Japan

\end{center}

\begin{abstract}
Two-dimensional Coulomb gases on an annulus at a special inverse 
temperature $\beta = 2$  are studied by using the orthogonal polynomial 
method borrowed from the theory of random matrices. The correlation 
functions among the Coulomb gas molecules are written in determinant 
forms and their asymptotic forms in the thermodynamic limit are evaluated. 
When the Coulomb gas system has a continuous rotational symmetry, the 
corresponding orthogonal polynomials are monomials, and one can see 
a universal behavior of the correlation functions in a thin annulus limit.
In a system with a discrete rotational symmetry, the corresponding orthogonal 
polynomials are not in general monomials, and a breakdown of the universality 
is observed.

\end{abstract}


\newpage

\section{Introduction}
\setcounter{equation}{0}
\renewcommand{\theequation}{1.\arabic{equation}}

The two-dimensional Coulomb gas system is one of the simplest models of 
interacting molecules. It has been a major subject of interest both in mathematics 
and physics \cite{pjf1, chafai}, because it sometimes gives valuable exactly solvable 
fluid models in more than one dimension, and has an application to physics of disordered and 
quantum systems, in particular by way of the theory of random matrices\cite{ginibre,pjf2}. 
\par
Suppose that $N$ classical Coulomb gas molecules are located at $(x_j,y_j)$ ($j=1,2,\cdots,N$) 
on the two-dimensional plane. Each molecule has a positive unit charge and logarithmically 
interacts with each other. The velocity-independent part of the 
gas molecule Hamiltonian is written in the form
\begin{equation}
{\cal H} = \sum_{j=1}^N {\cal V}(z_j) - \sum_{j<\ell}^N \log|z_j - z_\ell|
\end{equation}
with complex variables $z_j = x_j + i y_j$ ($j=1,2,\cdots,N$). We assume that the real-valued potential function ${\cal V}(z)$ 
satisfies ${\cal V}(z) =  {\cal V}({\bar z})$, where ${\bar z}$ is the complex conjugate of a complex number 
$z$. The probability density function at a thermal equilibrium with a temperature ${\cal T}$ is
\begin{equation}
\label{pdf}
P(z_1,z_2, \cdots,z_N) = \frac{1}{{\cal Z}_N}  e^{- \beta {\cal H}} = \frac{1}{{\cal Z}_N} \prod_{j=1}^N w(z_j) \prod_{j<\ell}^N  |z_j - z_\ell |^\beta, 
\end{equation}
and the associated integration measure is $d^2z_1 d^2z_2 \cdots d^2z_N$ ($d^2z_j = dx_j dy_j$). 
Here $\beta > 0$ is the inverse temperature $1/(k_B {\cal T})$ with the Boltzmann constant $k_B$, and $w(z) = 
e^{-\beta {\cal V}(z)}$ is called the weight function. In order to normalize the probability density function, the partition function
\begin{equation}
{\cal Z}_N = \int d^2z_1 \int d^2z_2 \cdots \int d^2z_N \prod_{j=1}^N w(z_j) \prod_{j<\ell}^N  |z_j - z_\ell |^\beta
\end{equation}
is included. The integration with respect to each variable $z_j$ is taken over the whole complex plane.
\par
In this paper we are interested in the $k$-molecule correlation functions
\begin{equation}
\label{correlation}
\rho(z_1,z_2,\cdots,z_k) = \frac{N!}{(N-k)!}  \int d^2z_{k+1} \int d^2z_{k+2} \cdots \int d^2z_N  P(z_1,z_2,\cdots,z_N). 
\end{equation}
The $k$-molecule correlation functions are typical physical quantities. In particular, the 
$1$-molecule correlation function $\rho(z)$ gives the molecule density. At a special inverse 
temperature $\beta = 2$, it is known in random matrix theory that the  correlation functions 
can be evaluated in determinant forms\cite{KS}
\begin{equation}
\rho(z_1,z_2,\cdots,z_k) = {\rm det}[K(z_j,z_\ell)]_{j,\ell = 1,2,\cdots,k}
\end{equation}
with
\begin{equation}
\label{kernel}
K(z_j,z_\ell) = \sqrt{w(z_j) w({\bar z_\ell})} \sum_{n=0}^{N-1} \frac{1}{h_n} p_n(z_j) p_n({\bar z_\ell}).
\end{equation}
Here $p_n(z)$ are corresponding orthogonal polynomials of $z$ satisfying the 
orthogonality relation
\begin{equation}
\label{orthogonality}
\int d^2z \ w(z) \ p_m({\bar z}) p_n(z) = h_n \delta_{mn}, \ \ \ m,n=0,1,2,\cdots,N-1,
\end{equation}
where $\delta_{mn}$ is the Kronecker delta. Note that $p_n({\bar z}) = \overline{p_n(z)}$ 
holds due to the conditions on the  weight  function $w(z)$. 
We moreover need to suppose that the orthogonal polynomial $p_n(z)$ is one 
of the following two types:
\par\medskip\noindent
Type A: $p_n(z)$ is a polynomial of degree $n$ with the highest degree term $z^n$.
\par\medskip\noindent
Type B: $p_n(z)$ is a polynomial of degree $N-1$ with the lowest degree term $z^n$.
\par\medskip\noindent
In both of Type A and Type B orthogonal polynomials $p_n(z)$, the coefficients of the terms except 
$z^n$ can be zero. 
\par
When the weight function $w(z)$ depends only on the radial parameter 
$r = |z|$ as $w(z) = f(r)$,  the Coulomb gas system is rotationally symmetric because of the invariance relation
\begin{equation}
\label{invariance}
w(z e^{i \varphi}) = w(z)
\end{equation}
for any real rotation angle $\varphi$. In such rotationally symmetric cases, one can see that the orthogonal 
polynomials $p_n(z)$ are monomials $z^n$. Monomials satisfy the orthogonality relation 
\begin{equation}
\int_0^{2 \pi} {\bar z}^m z^n d\theta = 0, \ \ \ m \neq n
\end{equation}
on a circle $z = r e^{i \theta}$ with a fixed radius $r > 0$.  Since this 
one-dimensional orthogonality relation on a circle holds for arbitrary $r$, it can be 
extended to the two-dimensional orthogonality on the complex plane as
\begin{equation}
\int d^2z \ w(z) \ {\bar z}^m z^n = \int_0^{\infty} dr \ r f(r) \ \int_0^{2 \pi} {\bar z}^m z^n d\theta = 0, \ \ \ m \neq n,
\end{equation}
if the integral over $r$ converges. Therefore the orthogonality relation (\ref{orthogonality}) 
holds for monomials $p_n(z) = z^n$ and a rotationally symmetric weight function $w(z) = f(r)$.
\par
One of the important applications of Coulomb gases is the theory of 
random matrices. In some typical non-hermitian random matrix models, the complex 
eigenvalues can be identified with Coulomb gas molecules in two dimensions. 
In particular, truncations of unitary matrices\cite{ZS}, gap probabilities and 
linear/counting statistics of complex eigenvalues\cite{APS,CL,charlier,BDM,BF}, 
induced ensembles\cite{FBKSZ} and products of random matrices\cite{AB} are related to 
rotationally symmetric Coulomb gases. Then 
we can analyze the eigenvalue distribution by means of monomials. In that respect, 
monomials have been powerful tools in the theory of non-hermitian random matrices.
\par
Szeg\H{o} gave a systematic mathematical argument to construct more general orthogonal 
polynomials\cite{szego}. Let us consider a circle 
\begin{equation}
U_r = \{ Z = X + i Y | X^2 + Y^2 = r^2\}
\end{equation}
($X,Y$ real) on the complex $Z$ plane with  a radius $r$. Suppose that $U_r$ is mapped 
to a closed curve $C_r$ on the complex $z$ plane by a one-to-one 
conformal map (Riemann map)  in the form 
\begin{equation}
z = \chi(Z) = \chi Z + \chi_0 +  \chi_1 Z^{-1} + \chi_2 Z^{-2} + \cdots
\end{equation}
with $\chi > 0$ and complex coefficients $\chi_j$ ($j = 0,1,2,\cdots$). 
\par
We assume that a complex valued function $D(z)$ for sufficiently large $|Z| = r$ (including $Z = \infty$) 
is regular and non-zero. Let us consider a necessary condition to satisfy the orthogonality relation 
\begin{equation}
\label{cr}
\int_{C_r} |D(z)|^2 p_m({\bar z}) p_n(z)  |dz| = 0, \ \ \ m \neq n
\end{equation}
for every sufficiently large $r$. Here $p_n(z)$ is a polynomial of $z$ with the highest 
degree term $z^n$ ($n=0,1,2,\cdots$), and 
\begin{equation}
|dz| = \sqrt{dx^2 + dy^2}
\end{equation}
with $z = x + i y$ ($x,y$ real) and $dr = 0$. Under the above conditions on $D(z)$, Szeg\H{o} proved 
that Riemann maps $\chi(Z)$, regular functions $D(z)$ and the corresponding (Type A) 
orthogonal polynomials $p_n(z)$ are restricted and essentially classified into five classes. 
A typical member of each class is listed below.
\par
\medskip
\noindent
(I) $\chi(Z) = Z$, $D(z)=1$, and $p_n(z) = z^n$.
\par
\medskip
\noindent
(II) $\chi(Z) = Z$, $D(z) = (1 - z^{-M})^{-1}$ with $|z| > 1$ and  a positive integer $M$, and 
\begin{equation}
p_n(z) = \left\{ \begin{array}{ll} z^n, & 0 \leq n < M, \\ 
z^{n - M} ( z^M - 1), & n \geq M. \end{array} \right.
\end{equation}
\par
\medskip
\noindent
(III) $\chi(Z) = (Z + Z^{-1})/2$, $D(z) = (1 - Z^{-2})^{-1/2}$ with $|Z| > 1$, and 
\begin{equation}
p_n(z) = p_n\left( \frac{Z+Z^{-1}}{2}  \right) = \left\{ \begin{array}{ll} 1, & n=0, \\ 
(Z^n + Z^{-n})/2^n, & n \geq 1. \end{array} \right. 
\end{equation}
\par
\medskip
\noindent
(IV) $\chi(Z) = (Z + Z^{-1})/2$, $D(z) = (1 - Z^{-2})^{1/2}$ with $|Z| > 1$, and
\begin{equation}
p_n(z) = p_n\left( \frac{Z+Z^{-1}}{2} \right) = \frac{Z^{n+1} - Z^{- n - 1}}{2^n (Z - Z^{-1})}.
\end{equation}
\par
\medskip
\noindent
(V) $\chi(Z) = (Z + Z^{-1})/2$, $D(z) = \sqrt{(1 + Z^{-1}) /(1 - Z^{-1})}$ with $|Z| > 1$, 
and 
\begin{equation}
p_n(z) = p_n\left( \frac{Z+Z^{-1}}{2} \right) = \frac{Z^{n+1} + Z^{-n}}{2^n (Z + 1)}.
\end{equation}
In the classes (I) and (II), $p_n(z)$ are orthogonal polynomials on a circle $|z| = r$. On the other 
hand, in the classes (III), (IV) and (V), $p_n(z)$ are the Chebyshev polynomials of the first, second and 
third kind. They are orthogonal polynomials on an ellipse\cite{szego, walsh1,walsh2}, because the Joukowsky transform 
$\chi(Z) = (Z+Z^{-1})/2$ maps the circle $|Z| = r>1$ to an ellipse. 
\par
If the orthogonality relation (\ref{cr}) holds for every sufficiently large $r$, the corresponding 
two-dimensional orthogonality relation 
\begin{equation}
\int f(r) dr \int_{C_r} |D(z)|^2 p_m({\bar z}) p_n(z) |dz| = 0, \ \ \  m \neq n
\end{equation}
holds for an arbitrary weight function $f(r)$. Here the integration over $r$ is taken 
in the region with sufficiently large $r$ and the convergence of the integral is 
assumed. Let us suppose that the relation between the measures is
\begin{equation}
dr |dz| = J(z) d^2z.
\end{equation}
Then we find 
\begin{equation}
\int  w(z)  p_m({\bar z}) p_n(z) d^2z = 0, \ \ \ m \neq n
\end{equation}
with 
\begin{equation}
w(z) = f(r) J(z) |D(z)|^2.
\end{equation}
For example, when $C_r$ is a circle $|z| = r$, we can set $z = r e^{i \theta}$ ($\theta$ real) and 
obtain $J(z) = 1$ and $w(z) = f(r) |D(z)|^2$. 
This formula can be used as the orthogonality relation (\ref{orthogonality}) on the complex plane to 
construct two-dimensional Coulomb gases. In \cite{elliptic}, the author studied such two-dimensional 
systems corresponding to the classes (III), (IV) and (V). This paper focuses on the classes (I) and (II), 
in particular the cases when the gas molecules are confined on an annulus. Coulomb gases on 
an annulus appear in the gap probability theory of random matrix eigenvalues\cite{charlier}. 
They are also physically interesting, because it has a thin annulus (quasi one-dimensional) limit, in which we can observe the transition between two-dimensional and one-dimensional gas systems.
\par
This paper is organized as follows. In \S 2, we consider the Coulomb gas systems in the class (I), 
in which the molecules are distributed on an annulus around the origin and each of them has 
a unit positive charge. As a result, in the limit of a thin annulus, universal forms of the correlation 
functions are obtained. In \S 3, the Coulomb gas systems in the class (II) are treated. Each of these 
systems has negative point charges fixed on the unit circle around the origin, and the annulus is put in the exterior of the unit circle. Due to the potential singularity caused by the negative 
charges, non-universal correlation functions are derived, when the annulus is in the vicinity of 
the unit circle. In \S 4, negative point charges are again fixed on the unit circle and the annulus 
is put in the interior of the unit circle.  Non-universal correlation functions are again observed, 
when the annulus is close to the unit circle. The last section is devoted to summary and 
discussion.

\section{Universal correlations on an annulus}
\setcounter{equation}{0}
\renewcommand{\theequation}{2.\arabic{equation}}

Let us first examine the thermodynamic limit $N \rightarrow \infty$ of the class (I) Coulomb gases 
with monomial  orthogonal polynomials $p_n(z) = z^n$. In this case, the kernel function $K(z_1,z_2)$ 
defined in (\ref{kernel}) is
\begin{equation}
\label{classI}
K(z_1,z_2) = \sqrt{f(r_1) f(r_2)} \sum_{n=0}^{N-1} \frac{1}{h_n} z_1^n  {\bar z_2}^n
\end{equation}
with
\begin{equation}
h_n = \int d^2z \ f(r) |z|^{2 n}.
\end{equation}
Here $r_1 = |z_1|$, $r_2= |z_2|$ and $r = |z|$.
\par
Now we assume that the Coulomb gas molecules are confined on an annulus
\begin{equation}
\label{annulus}
{\cal A} = \{z | R \leq |z| \leq v \}.
\end{equation}
That is, 
$f(r)$ has a form $f(r) = \Theta(r) g(r)$ with
\begin{equation}
\Theta(r) = \left\{ \begin{array}{ll} 1, & R \leq r \leq v, \\
0, & {\rm otherwise}. \end{array} \right.
\end{equation}
Here $R$ is the inner radius and $v$ is the outer radius of the annulus ($0 < R < v$). 
Let us next put a point charge $\Gamma$ at the origin. Note that $\Gamma$ can take an 
arbitrary real value. When a charge distribution in the region $|z| < R$ is rotationally symmetric around the 
origin, it can be replaced with such a point charge at the origin with no effect on the molecules on the annulus ${\cal A}$. 
Due to a charge $\Gamma$ at the origin, the potential function ${\cal V}(z)$ and the weight function 
$w(z) = f(r)$ are shifted as 
\begin{equation}
{\cal V}(z) \mapsto {\cal V}(z) - \Gamma \log |z|, \ \ \ f(r) \mapsto f(r) r^{2 \Gamma}.
\end{equation}
As this shift does not break the rotational invariance (\ref{invariance}), the orthogonal polynomials $p_n(z)$ are 
kept to be monomials $z^n$. On the annulus ($R \leq r_1,r_2 \leq v$) we consequently have
\begin{equation}
K(z_1,z_2) = \sqrt{g(r_1)  g(r_2)} (r_1 r_2)^{\Gamma}\sum_{n=0}^{N-1} \frac{1}{h_n} z_1^n  {\bar z_2}^n
\end{equation}
with
\begin{equation}
h_n =  2 \pi \int_R^v dr \ g(r) r^{2 n + 2 \Gamma +1}.
\end{equation}
\par
Let us introduce real angular variables $\theta_1,\theta_2$ as
\begin{equation}
\label{polar}
z_1 = r_1 e^{i \theta_1}, \ \ \ z_2 = r_2 e^{i \theta_2}
\end{equation}
($r_1,r_2 > 0$) and consider the thermodynamic limit $N \rightarrow \infty$.  
To begin with, we treat a two-dimensional Coulomb gas on a flat 
disc with $R \rightarrow 0$ and $g(r) = 1$. In order to have an 
normalizable molecule density in the limit$R \rightarrow 0$, we 
assume
\begin{equation}
\Gamma > -1.
\end{equation}
As
\begin{equation}
h_n =  2 \pi \int_0^v dr \ r^{2 n + 2 \Gamma +1} = \frac{2 \pi}{2 n + 2 \Gamma + 2} v^{2 n + 2 \Gamma + 2},
\end{equation}
one obtains
\begin{equation}
K(z_1,z_2) =  (r_1 r_2)^{\Gamma} \sum_{n = 0}^{N-1} \frac{1}{h_n} z_1^n {\bar z_2}^n 
= \frac{|\zeta|^{\Gamma}}{2 \pi v^2} \sum_{n=0}^{N-1} (2 n + 2 \Gamma + 2) \zeta^n,
\end{equation}
where $\zeta = z_1 {\bar z_2}/v^2$. Then we find
\begin{equation}
\label{dual-kI}
K(z_1,z_2) = \frac{|\zeta|^{\Gamma}}{\pi v^2 (1 - \zeta)} \left(
\frac{1 - \zeta^N}{1 - \zeta} - N \zeta^N  
+  \Gamma (1 - \zeta^N) \right).
\end{equation} 
Suppose that the molecules are apart from the disc edge, namely
\begin{equation}
|z_1| < v, \ \ \ |z_2| < v.
\end{equation}
Then it follows from $|\zeta| < 1$ that
\begin{equation}
\label{dual-kI-as}
K(z_1,z_2) \sim \frac{|\zeta|^{\Gamma}}{\pi v^2 (1- \zeta)} \left(
\frac{1}{1 - \zeta}  
+ \Gamma \right)
\end{equation}
in the limit $N \rightarrow \infty$ with $\zeta$ and $\Gamma$ fixed. 
\par
This asymptotic evaluation is not valid when the molecules are in the vicinity of the disc edge. In order 
to deal with the edge region, we introduce scaling variables $t_1,t_2,\phi_1$ and $\phi_2$ as
\begin{equation}
\label{scaling}
r_1 = v \left( 1 - \frac{t_1}{N} \right), \ \ \ r_2 = v \left( 1 - \frac{t_2}{N} \right), \ \ \ 
\theta_1 = \psi + \frac{\phi_1}{N}, \ \ \ \theta_2 = \psi + \frac{\phi_2}{N},
\end{equation}
where $\psi$ is a fixed angle ($0 \leq \psi < 2 \pi$). Moreover we suppose that $\Gamma$ 
can depend on $N$ and define another scaling parameter $\gamma$ as
\begin{equation}
\label{smallgamma}
\gamma = \lim_{N \rightarrow \infty} (\Gamma/N).
\end{equation}
Because of the condition $\Gamma > -1$, $\gamma \geq 0$ holds. The asymptotic form of the kernel function is derived as
\begin{eqnarray}
\label{kappa}
K(z_1,z_2) & = & \left( v \left( 1 - \frac{t_1}{N}  \right) v \left( 1 - \frac{t_2}{N} \right) 
\right)^{\Gamma} \sum_{n=0}^{N-1} \frac{2 n + 2 \Gamma + 2}{2 \pi v^{2 n + 2 \Gamma + 2}} 
\nonumber \\ 
& & \times  v^{2 n} \left( 1 - \frac{t_1}{N} \right)^n \left( 1 - \frac{t_2}{N} \right)^n 
e^{i (n/N) (\phi_1 - \phi_2)} \nonumber \\ 
& \sim & \kappa(z_1,z_2) = \frac{N^2}{\pi v^2} \int_0^1 dc \ (c + \gamma) e^{-(c + \gamma) (t_1 + t_2)} 
e^{i c (\phi_1 - \phi_2)} \nonumber \\ 
& = & \frac{N^2}{\pi v^2} \frac{e^{-(\tau + i \phi) \gamma}}{\tau} 
\left( \gamma - (1 + \gamma) e^{-\tau} + \frac{1 - e^{-\tau}}{\tau} \right)
\end{eqnarray}
in the limit $N \rightarrow \infty$. Here 
$\phi=\phi_1 - \phi_2$ and $\tau = t_1 + t_2 - i \phi$. This asymptotic form is known 
in the study of a sub-block of a random unitary matrix (Eq. (3.2.122) of \cite{fischmann}, 
see also \cite{BF}). In particular, a special case $\gamma=0$ is a well-known 
universal result\cite{ZS,gluck-PRL,gluck-PRE,FS,lubinsky,CW}. 
The asymptotic molecule density 
$\kappa(z,z)$ satisfies
\begin{equation}
\int \kappa(z,z) d^2z = \frac{N}{\pi} \int_0^{2 \pi} d\theta \int_0^{\infty} dt \ 
\frac{e^{- 2 t \gamma}}{2 t} \left( \gamma - (1 + \gamma) e^{- 2 t} + 
\frac{1 - e^{-2 t}}{2 t} \right)= N,
\end{equation}
which means that almost all molecules are accumulated in the interior vicinity of the disc edge, 
as reported in \cite{ACC} for general Coulomb gases with hard walls.
\par
Let us next consider a Coulomb gas in the exterior region of a disc with a radius 
$R >0$. In order to see that case, we go back to the annulus with $0 < R < v$ and 
take the limit $v \rightarrow \infty$. The one-body potential function is again assumed 
to be flat ($g(r) = 1$). Under the assumption 
\begin{equation}
\Gamma < -N
\end{equation}
and the notation (\ref{polar}), we find
\begin{equation}
h_n =  2 \pi \int_R^\infty dr \ r^{2 n + 2 \Gamma +1} = - \frac{2 \pi}{2 n + 2 \Gamma + 2}  
R^{2 n + 2 \Gamma + 2}
\end{equation}
and
\begin{eqnarray}
\label{dual-kII}
K(z_1,z_2) & = &  (r_1 r_2)^{\Gamma} \sum_{n = 0}^{N-1} \frac{1}{h_n} z_1^n {\bar z_2}^n 
= - \frac{|\eta|^{\Gamma}}{2 \pi R^2} \sum_{n=0}^{N-1} (2 n + 2 \Gamma + 2) \eta^n 
\nonumber \\
& = & \frac{|\eta|^{\Gamma}}{\pi R^2 (\eta-1)} \left(
\frac{\eta^N-1}{\eta-1} - N \eta^N  - \Gamma (\eta^N-1) \right),
\end{eqnarray} 
where $\eta = z_1 {\bar z_2}/R^2$. If the molecules are apart from the disc edge, which means
\begin{equation}
|z_1| > R, \ \ \ |z_2| > R,
\end{equation}
one can see that $|\eta| > 1$ and
\begin{equation}
K(z_1,z_2) \sim \frac{|\eta|^{-{\tilde \Gamma}}}{\pi R^2 (\eta-1)} \left(
\frac{1}{\eta-1}  
+ {\tilde \Gamma} \right) e^{i N (\theta_1 - \theta_2)} 
\end{equation}
in the limit $N \rightarrow \infty$ with $\eta$ and ${\tilde \Gamma}=-N-\Gamma$ fixed. Note that the factor $e^{i N (\theta_1 - \theta_2)} $ can be replaced with $1$ without changing the correlation functions.
\par
By using mappings 
\begin{equation}
z_1 \mapsto \frac{1}{z_1}, \ \ \ z_2 \mapsto \frac{1}{z_2}, \ \ \ R \mapsto \frac{1}{v}
\end{equation}
and a parameter mapping
\begin{equation}
\Gamma \mapsto - \Gamma - N - 1
\end{equation}
coming from (\ref{bridge}), $|z_1 z_2|^2 K(z_1,z_2)$ (with $K(z_1,z_2)$ in (\ref{dual-kII})) is mapped 
to $(\zeta/|\zeta|)^{-N+1} K(z_1,z_2)$ (with $K(z_1,z_2)$ in (\ref{dual-kI})). 
Note that the extra phase factor $(\zeta/|\zeta|)^{-N+1}$ does not change 
the correlation functions. This is anticipated from the duality relation 
(\ref{duality-relation}) in Appendix A, and thus the similarity of (\ref{dual-kI}) and 
(\ref{dual-kII}) is explained. 
\par
In order to treat the molecules in the exterior vicinity of the disc edge, we define 
$\sigma_1,\sigma_2,\phi_1$ and $\phi_2$ as
\begin{equation}
r_1 = R \left( 1 + \frac{\sigma_1}{N} \right), \ \ \ r_2 = R \left( 1 +\frac{\sigma_2}{N} \right), \ \ \ 
\theta_1 = \psi + \frac{\phi_1}{N}, \ \ \ \theta_2 = \psi + \frac{\phi_2}{N},
\end{equation}
where $\psi$ is again a fixed angle ($0 \leq \psi < 2 \pi$). The scaling parameter $\gamma$ 
defined in (\ref{smallgamma}) satisfies $\gamma \leq -1$ due to the condition $\Gamma < -N$.
As before we can derive the asymptotic form of the kernel function as
\begin{eqnarray}
\label{tilde-kappa}
K(z_1,z_2)  & \sim & {\tilde \kappa}(z_1,z_2) = - \frac{N^2}{\pi R^2} \int_0^1 dc \ (c + \gamma) e^{(c + \gamma) (\sigma_1 + \sigma_2)} 
e^{i c (\phi_1 - \phi_2)} \nonumber \\ 
& = & \frac{N^2}{\pi R^2} \frac{e^{(\sigma - i \phi) \gamma}}{\sigma} 
\left( \gamma - (1 + \gamma) e^{\sigma} + \frac{e^{\sigma}-1}{\sigma} \right)
\end{eqnarray}
in the limit $N \rightarrow \infty$. Here 
$\phi=\phi_1 - \phi_2$ and $\sigma = \sigma_1 + \sigma_2 + i \phi$. As
\begin{equation}
\int {\tilde \kappa}(z.z) d^2z = \frac{N}{\pi} \int_0^{2 \pi} d\theta \int_0^{\infty} dt \ 
\frac{e^{2 t \gamma}}{2 t} \left( \gamma - (1 + \gamma) e^{2 t} + 
\frac{e^{2 t}-1}{2 t} \right)= N,
\end{equation}
one can say that almost all molecules are accumulated in the exterior vicinity of the disc edge, 
because of the attraction from the negative charge at the origin.
\par 
In order to explain the similarity of $\kappa(z_1,z_2)$ and ${\tilde \kappa}(z_1,z_2)$ originating from 
the duality relation (\ref{duality-relation}), we employ mappings 
\begin{equation}
z_1 \mapsto \frac{1}{z_1}, \ \ \ z_2 \mapsto \frac{1}{z_2}, \ \ \ R \mapsto \frac{1}{v}
\end{equation}
leading to
\begin{equation}
\psi \mapsto -\psi, \ \ \ \phi \mapsto -\phi, \ \ \ \sigma \mapsto \tau 
\end{equation}
together with
\begin{equation}
\gamma \mapsto - \gamma - 1
\end{equation}
in the limit $N \rightarrow \infty$. Then $|z_1 z_2|^2 {\tilde \kappa}(z_1,z_2)$ (with ${\tilde \kappa}(z_1,z_2)$ in (\ref{tilde-kappa})) is mapped to $e^{-i \phi} \kappa(z_1,z_2)$ (with $\kappa(z_1,z_2)$ in (\ref{kappa})), as expected. Here again the extra phase factor $e^{-i \phi}$ does not change the correlation functions. 
\par
Now we again go back to the annulus with $0 < R < v$ with the weight function $g(r)$ and suppose 
that $\Gamma$ can take an arbitrary real value. In order to study the transition from an 
annulus to a circle, we focus on the thin annulus limit in which the inner edge is in the vicinity of the outer edge. For that purpose one introduces real scaling parameters $T$ and $t$ as
\begin{equation}
\label{scaling-T-t}
R = v \left(1 - \frac{T}{N} \right), \ \ \ r = v \left( 1 - \frac{t}{N} \right)
\end{equation}
with $0<t < T$. The universal forms of the correlation functions are obtained 
in the limit $N \rightarrow \infty$ in terms of the scaling variables $t_1,t_2,\phi_1$ and 
$\phi_2$ defined in (\ref{scaling}) with $0 < t_1,t_2 < T$.

Under the condition that $g(r)$ is bounded and continuous for $r \leq v$, one 
is able to find an estimate
\begin{eqnarray}
h_n & =  & 
2 \pi \int_{v \left( 1  - \frac{T}{N} \right)}^v dr \ g(r) r^{2 n+2 \Gamma + 1} 
\nonumber \\ 
& = & \frac{2 \pi v}{N} \int_0^T dt \ g\left( v \left( 1 - \frac{t}{N} \right) \right) 
\left( v  \left( 1 - \frac{t}{N} \right) \right)^{2 n+2 \Gamma+1} \nonumber \\ 
& \sim & \frac{2 \pi v^{2  n + 2 \Gamma + 2}}{N} g(v) \int_0^T dt \ 
e^{- 2 (c + \gamma) t} = \frac{\pi v^{2 n + 2 \Gamma + 2}}{N (c + \gamma)}  
g(v) (1 - e^{- 2 (c + \gamma) T} ) \nonumber \\ 
\end{eqnarray}
in the limit $N \rightarrow \infty$ with $c = \lim_{N \rightarrow \infty} (n/N)$ fixed.  
We additionally assume that $g(v) > 0$. Then the asymptotic limit of the kernel function is
\begin{eqnarray}
\label{universal}
& & K(z_1,z_2) \nonumber \\ 
& = & \sqrt{g\left(v \left( 1 - \frac{t_1}{N} \right) \right) g\left(v \left( 1 - \frac{t_2}{N} \right) \right)} 
\left( v \left( 1 - \frac{t_1}{N} \right) v \left( 1 - \frac{t_2}{N} \right)  \right)^\Gamma \nonumber \\ 
& & \times
\sum_{n=0}^{N-1} \frac{v^{2 n}}{h_n} 
\left(1 - \frac{t_1}{N} \right)^n  
\left(1 - \frac{t_2}{N} \right)^n e^{i (n/N) (\phi_1 - \phi_2)} \nonumber \\ 
& \sim & \frac{N^2}{\pi v^2} \int_0^1 dc \ (c + \gamma) \ \frac{e^{- (c + \gamma) (t_1 + t_2)} 
e^{i c (\phi_1 - \phi_2)}}{1 - e^{- 2 (c + \gamma) T}}, \ \ \ 
N \rightarrow \infty.
\end{eqnarray}
Let us remark that this asymptotic limit does not depend on the specific form of $g(r)$.  
It was reported in a special case $\Gamma = 0$ and $g(r) = r^\alpha$ ($\alpha$ is real and fixed)  
in \cite{elliptic}, and 
now it turned out to hold for more general $\Gamma$ and $g(r)$. 
Moreover a similar limit can also be derived for Coulomb gas models on an elliptic annulus, if appropriate 
scalings are adopted\cite{elliptic}. Thus we claim that 
it gives a universal form of the Coulomb gas correlation functions.
\par
In the subsequent one-dimensional limit $T \rightarrow 0$, we also need to take the limit $t_1,t_2 \rightarrow 0$ and a Coulomb gas 
on the circle $|z| = v$ is obtained. The asymptotic form of the kernel function is
\begin{equation}
\label{sine-kernel}
K(z_1,z_2) \sim \frac{N^2}{2 \pi v^2 T} \int_0^1 \ dc \ e^{i c (\phi_1 - \phi_2)} = \frac{N^2}{2 \pi v^2 T} e^{i(\phi_1 - \phi_2)/2} 
\frac{\sin((\phi_1 - \phi_2)/2)}{(\phi_1 - \phi_2)/2}.
\end{equation}
This is called the sine kernel and a well-known universal result in random matrix theory\cite{dyson,mehta}. 
It appears here, because the Coulomb gas on the unit circle gives the eigenvalue distribution of random unitary matrices. As the phase 
factor $e^{i(\phi_1 - \phi_2)/2}$ can be removed without changing the determinant forms, we can see that 
the asymptotic forms of the correlation functions are
\begin{equation}
\rho(z_1,z_2,\cdots,z_k) \sim (\rho_0)^k {\rm det}\left[\frac{\sin((\phi_j - \phi_\ell)/2)}{(\phi_j - \phi_\ell)/2} 
\right]_{j,\ell = 1,2,\cdots,k},
\end{equation}
where $\rho_0 = N^2/(2 \pi v^2 T)$ is the one-molecule density. Note that the correlation functions do not 
depend on the parameter $\gamma$ at all. Thus one can say that the Coulomb gas distribution on the circle $|z|=v$ 
is not affected by the point charge located at the origin. In general it is not affected by charges in 
the interior region $|z| < v$, if the charge distribution has a rotational symmetry around the origin.
\par
In order to study the opposite limit $T \rightarrow \infty$, we need to separately consider the 
following three cases and utilize an asymptotic relation
\begin{equation}
1 - e^{- 2 (c + \gamma) T} \sim \left\{ \begin{array}{ll} 1, & c + \gamma > 0, \\ 
 - e^{-2(c + \gamma) T}, &  c + \gamma < 0. \end{array} \right. 
\end{equation}
\par
\medskip
\noindent
(1) $\gamma \geq 0$
\par
\medskip
\noindent
In this case, due to the repulsion among the molecules and 
the positive charge at the origin, the molecules are accumulated 
in the vicinity of the outer boundary of the annulus. Taking the limit 
$T \rightarrow \infty$ with the parameters $\phi=\phi_1 - \phi_2$ 
and $\tau = t_1 + t_2 - i \phi$ fixed, we obtain
\begin{equation}
\label{kappa1}
K(z_1,z_2) \sim \kappa_1(z_1,z_2) = \frac{N^2}{\pi v^2} \frac{e^{- (\tau + i \phi) \gamma}}{\tau} 
\left( \gamma - (1 + \gamma) e^{- \tau} + \frac{1 - e^{-\tau}}{\tau} \right).
\end{equation} 
This is identical to the asymptotic limit (\ref{kappa}) of the kernel function at the 
disc edge, as expected. As before, since $\displaystyle \int \kappa_1(z.z) d^2z = N$, we 
can say that almost all molecules are accumulated in the vicinity of the outer boundary.
\par
\medskip
\noindent
(2) $-1 < \gamma <0$
\par
\medskip
\noindent
In this intermediate case the negative charge at the origin and the positive charges on the 
molecules are balanced. The molecules are accumulated in the regions close to both outer 
and inner boundaries. In the vicinity of the outer boundary, we have 
\begin{equation}
\label{kappa2}
K(z_1,z_2) \sim \kappa_2(z_1,z_2) = \frac{N^2}{\pi v^2} \frac{e^{-  i \phi \gamma}}{\tau} 
\left(  \frac{1 - e^{-\tau (1 + \gamma)}}{\tau} - (1 + \gamma) e^{-\tau (1 + \gamma)} \right)
\end{equation} 
in the limit $T \rightarrow \infty$ with $\tau$ fixed. As we find
\begin{eqnarray}
\int \kappa_2(z,z) d^2z & = & \frac{N}{\pi} \int_0^{2 \pi} d\theta \int_0^{\infty} dt \ 
\frac{1}{2 t} \left( \frac{1 - e^{-2 t(1 + \gamma) }}{2 t} - (1 + \gamma) e^{- 2 t (1+ \gamma)}  \right) 
\nonumber \\ & = & N (1 + \gamma),
\end{eqnarray}
it can be said that most of $N(1 + \gamma)$ molecules are accumulated in the vicinity of the 
outer boundary.
\par
On the other hand, in order to see the vicinity of the inner
boundary, let us introduce a parameter 
$\sigma = 2 T - \tau = 2 T - t_1 - t_2 + i \phi$ and consider the limit $T \rightarrow \infty$ 
with $\sigma$ fixed. Then we find
\begin{equation}
\label{kappa2tilde}
K(z_1,z_2) \sim {\tilde \kappa}_2(z_1,z_2) = \frac{N^2}{\pi v^2} \frac{e^{-  i \phi \gamma}}{\sigma} 
\left(  \frac{1 - e^{\sigma \gamma}}{\sigma} + \gamma e^{\sigma \gamma} \right),
\end{equation}
where $\displaystyle \int {\tilde \kappa}_2(z,z) d^2z = - N \gamma$. That is, most of $- N \gamma$ 
molecules are accumulated in the vicinity of the inner boundary. 
\par
In order to see the effect of the duality relation (\ref{duality-relation}), we again 
introduce mappings
\begin{equation}
z_1 \mapsto \frac{1}{z_1}, \ \ \ z_2 \mapsto \frac{1}{z_2}, \ \ \ v \mapsto \frac{1}{v}.
\end{equation}
Then, in the limit $N \rightarrow \infty$, one obtains mappings
\begin{equation}
\psi \mapsto -\psi, \ \ \ \phi \mapsto -\phi, \ \ \ \sigma \mapsto \tau 
\end{equation}
and
\begin{equation}
\gamma \mapsto - \gamma - 1,
\end{equation}
so that $|z_1 z_2|^2  {\tilde \kappa}_2(z_1,z_2)$ is mapped to $e^{-i \phi} \kappa_2(z_1,z_2)$, where the phase 
factor $e^{-i \phi}$ does not affect the correlation functions. This result gives the duality between 
the inner and outer boundaries of the thin annulus.
\par
\medskip
\noindent
(3) $\gamma \leq -1$
\par
\medskip
\noindent
In this case, because of the attraction from the negative charge 
at the origin, the molecules are accumulated in the vicinity 
of the inner boundary. We can see that
\begin{equation}
\label{kappa3}
K(z_1,z_2) \sim \kappa_3(z_1,z_2) = \frac{N^2}{\pi v^2} \frac{e^{(\sigma - i \phi) \gamma}}{\sigma} 
\left(  \gamma -(1 + \gamma) e^{\sigma} + \frac{e^{\sigma} - 1}{\sigma}  \right)
\end{equation} 
in the limit $T \rightarrow \infty$ with $\sigma$ fixed. As expected, this is identical to the asymptotic limit (\ref{tilde-kappa}), when $v$ is replaced with $R$. Here $\displaystyle 
\int \kappa_3(z,z) d^2z = N$, which implies that almost all molecules are accumulated 
in the vicinity of the inner boundary.
\par
\medskip
In the analysis of the thin annulus limit, we have assumed that the weight function 
$g(r)$ satisfies $g(v) > 0$. In an attempt to see the necessity of this assumption, we 
examine a special case $v=1$ with a weight function
\begin{equation}
g(r) = (1 - r^2)^{\nu-1},
\end{equation}
where $\nu$ is a positive integer. This model is related to the eigenvalue distribution of truncated 
unitary matrices\cite{ZS,BF,fischmann}, as later discussed in \S 5. 
The assumptions $g(1) > 0$ is not satisfied when $\nu \neq 1$. In order to analyze the thin 
annulus limit, the scaling variable $T$ is defined as $R = 1 - (T/N)$. Then in the 
limit $N \rightarrow \infty$ we obtain
\begin{equation}
h_n = 2 \pi \int_R^1 dr \ g(r) \ r^{2 n + 2 \Gamma + 1} \sim \pi \left( \frac{2}{N} \right)^\nu \int_0^T dt \ t^{\nu-1} e^{-2 (c + \gamma) t}
\end{equation}
and
\begin{eqnarray}
K(z_1,z_2) & = & \sqrt{g(r_1) g(r_2)} (r_1 r_2)^{\Gamma} \sum_{n=0}^{N-1} \frac{1}{h_n} z_1^n {\bar z_2}^n 
\nonumber \\ 
& \sim & \frac{N^2}{2 \pi} (t_1 t_2)^{(\nu-1)/2} \int_0^1 dc \ 
\frac{e^{-(c + \gamma)(t_1 + t_2)} e^{i c (\phi_1 - \phi_2)}}{
\displaystyle \int_0^T dt \ t^{\nu-1} e^{-2 (c + \gamma) t}},
\end{eqnarray}
where polar coordinates (\ref{polar})  and scaling variables (\ref{scaling}) with $v=1$ are used. Except in the case $\nu=1$, this result is different from the universal form (\ref{universal}). We have thus seen 
that the assumption $g(1)>0$ is necessary.

\section{Non-universal correlations on an annulus}
\setcounter{equation}{0}
\renewcommand{\theequation}{3.\arabic{equation}}

Let us next study the class (II) model on an annulus in order to observe a breakdown of the universality. Suppose that the gas molecules are distributed on an annulus with an inner radius $R$ and an outer radius $v$ 
($1 < R < v$). As in previous section, a point charge $\Gamma$ is put at the origin. Moreover we put 
a negative unit charge at each of the regular polygon vertices $z = \omega^j$ ($j=0,1,2,\cdots,M-1$) 
with $\omega = e^{2 \pi i/M}$ ($M$ is a positive integer). These negative charges on the unit circle $|z| =1$ 
shift  the potential function ${\cal V}(z) - \Gamma \log|z|$ and the weight function $w(z) = f(r) r^{2 \Gamma}$ as
\begin{eqnarray}
\label{potential-and-weight}
{\cal V}(z) & \mapsto &  {\cal V}(z) + \sum_{j=0}^{M-1} \log |z - \omega^j| = {\cal V}(z) + \log|z^M - 1|, 
\nonumber \\
f(r) & \mapsto & f(r) |z^M - 1|^{-2} = f(r) r^{-2 M} |D(z)|^2,
\end{eqnarray}
where
\begin{equation}
D(z) = \frac{1}{1 - z^{-M}}.
\end{equation}
This Coulomb gas system no longer has a continuous rotational symmetry, 
Instead, as $D(z)$ is invariant under the discrete rotations
\begin{equation}
z \rightarrow z \omega^j, \ \ \ \omega = e^{2 \pi i/M}, \ \ \ j = 0,1,2,\cdots,M-1
\end{equation}
around the origin, it has a discrete rotational symmetry. 
\par
The kernel function defined in (\ref{kernel}) on the annulus has the form
\begin{equation}
\label{kernel-annulus}
K(z_1,z_2) = \sqrt{g(r_1)  g(r_2)} (r_1 r_2)^{\Gamma-M} |D(z_1) D(z_2)| \sum_{n=0}^{N-1} \frac{1}{h_n} p_n(z_1) p_n( {\bar z_2})
\end{equation}
with $1 < R \leq r_1,r_2 \leq v$ ($r_1 = |z_1|$, $r_2 = |z_2|$). Here the Type A orthogonal polynomials $p_n(z)$ with the highest degree term $z^n$ satisfy the orthogonality relation
\begin{eqnarray}
& & \int d^2z \ f(r) \ r^{2 (\Gamma - M)} \  |D(z)|^2 p_m({\bar z}) p_n(z)  \nonumber \\
& = & \int_R^v dr  \int_0^{2 \pi} d\theta \ r^{2(\Gamma - M)+1} \ g(r) \ |D(z)|^2 \ p_m(\bar{z})  p_n(z) = h_n \delta_{mn},
\end{eqnarray}
where $z = r e^{i \theta}$ ($r = |z| >1$). It follows from the orthogonality relation (\ref{A-orthogonality}) in Appendix B that 
$p_n(z)$ are given by
\begin{equation}
p_n(z) = \left\{ \begin{array}{ll} z^n, & 0 \leq n < M, \\ 
z^{n - M} ( z^M - 1), & n \geq M \end{array} \right.
\end{equation}
with
\begin{equation}
\label{hn}
h_n = \left\{ \begin{array}{ll} \displaystyle 
2 \pi \int_R^v dr \ g(r) \frac{r^{2 (n + \Gamma - M)+1}}{1 - r^{- 2 M}}, & 0 \leq n < M,  \\ 
\displaystyle 2 \pi \int_R^v dr \ g(r) r^{2 (n + \Gamma - M)+1}, & n \geq M.  \end{array} \right.
\end{equation}
Then, in the case $M < N$, we can separate $K(z_1,z_2)$ into two parts as
\begin{equation}
K(z_1,z_2) = K^{(1)}(z_1,z_2) + K^{(2)}(z_1,z_2),
\end{equation}
where
\begin{equation}
K^{(1)}(z_1,z_2) = \frac{\sqrt{g(r_1)  g(r_2)} (r_1 r_2)^\Gamma}{|z_1^M - 1| | z_2^M - 1|} 
\sum_{n=0}^{M-1} \frac{(z_1 {\bar z_2})^n}{h_n}
\end{equation}
and
\begin{equation}
K^{(2)}(z_1,z_2) = \frac{\sqrt{g(r_1)  g(r_2)} (r_1 r_2)^\Gamma}{|z_1^M - 1| | z_2^M - 1|} \sum_{n=M}^{N-1} \frac{1}{h_n} 
z_1^{n - M} (z_1^M- 1) {\bar z_2}^{n - M} ({\bar z_2}^M - 1). 
\end{equation}

\subsection{Fixed number of negative charges}

When the number $M$ of the negative charges is fixed, let us consider the thin annulus 
limit $N \rightarrow \infty$ of $K(z_1,z_2)$ with $T$ and $t$ fixed. Here $T$ and $t$ are 
scaling variables defined in (\ref{scaling-T-t}) with $v > 1$. As before, $g(r)$ is 
bounded and continuous for $r \leq v$, and $g(v) > 0$. When $n < M$, it can readily be seen that
\begin{equation}
h_n = 2 \pi \int_R^v dr \ g(r) \frac{r^{2 (n + \Gamma - M)+1}}{1 - r^{- 2 M}} \sim 
\frac{\pi v^{2(n + \Gamma - M) + 2} g(v)}{N \gamma} \frac{1 - e^{-2 \gamma T}}{1 - v^{- 2 M}}, \ \ \ N \rightarrow \infty.
\end{equation}
When $n \geq M$, on the other hand, we find
\begin{equation}
h_n = 2 \pi \int_R^v dr \ g(r) r^{2(n + \Gamma - M)+1} \sim 
\frac{\pi v^{2 (n + \Gamma - M)+2}}{N (c + \gamma)} g(v) (1 - e^{- 2 (c+\gamma) T}), \ \ \ N \rightarrow \infty.
\end{equation}
Here $c = \lim_{N \rightarrow \infty} (n/N) $ and $\gamma = \lim_{N \rightarrow \infty} (\Gamma/N)$ are fixed.  Introducing the polar coordinates as
\begin{equation}
z_1 = r_1 e^{i \theta_1}, \ \ \ z_2 = r_2 e^{i \theta_2},
\end{equation}
and using the scaling variables $t_1$, $t_2$, $\phi_1$ and $\phi_2$ defined in (\ref{scaling}), we are able to derive the asymptotic relations
\begin{equation}
K^{(1)}(z_1,z_2) \sim \frac{NM}{2 \pi v^2} \frac{2 \gamma}{1 - e^{-2 \gamma T}}
\frac{e^{-\gamma (t_1 + t_2)} (v^{2 M} - 1)}{v^{2 M} + 1 - 2 v^M \cos(M \psi)}, 
\ \ \ N \rightarrow \infty
\end{equation}
and
\begin{equation}
K^{(2)}(z_1,z_2) \sim \frac{N^2}{\pi v^2} \int_0^1 dc \ (c + \gamma) \frac{e^{-(c + \gamma) (t_1 + t_2)} e^{i c (\phi_1 - \phi_2)}}
{1 - e^{-2 (c + \gamma) T}}, \ \ \ N \rightarrow \infty.
\end{equation}
Since $K^{(1)}(z_1,z_2)$ is negligible compared to  $K^{(2)}(z_1,z_2)$ in the limit $N \rightarrow \infty$, the kernel function $K(z_1,z_2)$ can be evaluated as
\begin{equation}
K(z_1,z_2) \sim \frac{N^2}{\pi v^2} \int_0^1 dc \ (c + \gamma) \frac{e^{-(c + \gamma) (t_1 + t_2)} e^{i c (\phi_1 - \phi_2)}}
{1 - e^{-2 (c + \gamma) T}}, \ \ \ N \rightarrow \infty.
\end{equation}
This is identical to the universal formula (\ref{universal}). Thus one sees that the universality still holds under the presence of 
negative charges on the unit circle $|z|=1$. 
\par
The breakdown of the universality is observed when the thin annulus is in the outer neighborhood of the unit circle. In that case we introduce real scaling parameters $u$ and $T$ as
\begin{equation}
\label{scaling-T-u}
v = 1 + \frac{u}{N}, \ \ \ R = v - \frac{T}{N} = 1 + \frac{u - T}{N}
\end{equation}
with $0 < T < u$. Here $g(r)$ is supposed to be bounded and continuous for $r \geq 1$, 
and $g(1) > 0$. When $n < M$, it follows from (\ref{hn}) that 
\begin{eqnarray} 
h_n & = & 2 \pi \int_R^v dr \ g(r) \frac{r^{2 (n + \Gamma - M) + 1}}{1 - r^{-2 M}} 
\nonumber \\ 
& = & 
\frac{2 \pi}{N} \int_{u-T}^u dt \ g\left( 1 + \frac{t}{N} \right) 
\frac{\displaystyle \left(1 + \frac{t}{N} \right)^{2 (n + \Gamma - M)+1}}{\displaystyle 
1 - \left(1 + \frac{t}{N} \right)^{-2 M}} 
\nonumber \\ 
& \sim & \frac{\pi g(1)}{M} \int_{u - T}^u dt \ \frac{e^{2 \gamma t}}{t} = 
\frac{\pi g(1) e^{2 \gamma u}}{M} \int_0^T dt \ \frac{e^{-2 \gamma t}}{u -t}, \ \ \ N \rightarrow \infty.
\end{eqnarray}
When $n \geq M$, on the other hand, we obtain
\begin{eqnarray} 
h_n & = & 2 \pi \int_R^v dr \ g(r) r^{2 (n + \Gamma - M) + 1}
\nonumber \\ 
& = & 
\frac{2 \pi}{N} \int_{u-T}^u dt \ g\left( 1 + \frac{t}{N} \right) 
\left(1 + \frac{t}{N} \right)^{2 (n + \Gamma - M)+1}
\nonumber \\ 
& \sim & \frac{2 \pi g(1)}{N} \int_{u - T}^u dt \ e^{2 (c + \gamma) t} \nonumber \\ 
& = & \frac{\pi g(1)}{N(c + \gamma)} e^{2 (c + \gamma)u} \left( 1 - e^{- 2 (c + \gamma) T} \right), \ \ \ 
N \rightarrow \infty.
\end{eqnarray} 
Corresponding to the polar coordinates $r_1,r_2,\theta_1$ and $\theta_2$ defined as
\begin{equation}
z_1 = r_1 e^{i \theta_1}, \ \ \ z_2 = r_2 e^{i \theta_2},
\end{equation}
we introduce real scaling variables $t_1,t_2,\phi_1$ and $\phi_2$ as
\begin{equation}
\label{scaling1}
r_1 = v -  \frac{t_1}{N} = 1 + \frac{u - t_1}{N}, \ \  
r_2 = v -  \frac{t_2}{N} = 1 + \frac{u - t_2}{N}, \ \ 
\theta_1 = \psi + \frac{\phi_1}{N}, \ \ 
\theta_2 = \psi + \frac{\phi_2}{N},
\end{equation}
where $0 < t_1,t_2 < T$ and $\psi$ is a real fixed angle. Noting the asymptotic 
relations
\begin{equation}
z_j^M - 1 \sim \left\{ 
\begin{array}{ll} e^{i M \psi} - 1, & e^{i M \psi} \neq 1, \\
\displaystyle \frac{M}{N} s_j, & e^{i M \psi} = 1 \end{array} \right.
\end{equation}
in the limit $N \rightarrow \infty$ with $s_j = u - t_j + i \phi_j$ ($j = 1,2$), 
we find 
\begin{equation}
\label{non-universal-k1}
K^{(1)}(z_1,z_2) \sim \left\{ \begin{array}{ll} \displaystyle 
\frac{M^2 e^{ - \gamma (t_1 + t_2)}}{2 \pi ( 1 - \cos M \psi)} 
\frac{1}{\displaystyle \int_0^T dt \ 
\frac{e^{-2 \gamma t}}{u-t}}, & e^{i M \psi} \neq 1, \\ 
\displaystyle 
 \frac{N^2 e^{-\gamma (t_1 + t_2)}}{\pi |s_1 s_2|} \frac{1}{\displaystyle \int_0^T dt 
 \ \frac{e^{-2 \gamma t}}{u-t} }, & e^{i M \psi} = 1 \end{array} \right.
\end{equation}
and
\begin{equation}
\label{non-universal-k2}
K^{(2)}(z_1,z_2) \sim \left\{ \begin{array}{ll} \displaystyle 
\frac{N^2}{\pi}   \int_0^1 dc \ (c +\gamma)
\frac{e^{-(c +\gamma) (t_1 + t_2)} e^{i c (\phi_1 - \phi_2)}}{1 - e^{- 2 (c + \gamma) T}},  & e^{i M \psi} \neq 1, \\ 
\displaystyle 
\frac{N^2}{\pi} \frac{s_1 {\bar s}_2}{|s_1 s_2|} \int_0^1 d c \ (c + \gamma)  \frac{e^{-(c +\gamma) (t_1 + t_2)} e^{i c (\phi_1 - \phi_2)}}{1 - e^{- 2 (c + \gamma) T}}, & e^{i M \psi} = 1 \end{array} \right.
\end{equation}
in the limit $N \rightarrow \infty$. Putting the above asymptotic formulas together, we finally arrive 
at the universal formula 
((\ref{universal}) with $v = 1$)
\begin{equation}
\label{universal-Mfixed}
K(z_1,z_2) \sim \frac{N^2}{\pi}  \int_0^1 dc \ (c +\gamma)
\frac{e^{-(c +\gamma) (t_1 + t_2)} e^{i c (\phi_1 - \phi_2)}}{1 - e^{- 2 (c + \gamma) T}}, 
\ \ \ N \rightarrow \infty
\end{equation}
for $e^{i M \psi} \neq 1$, and a non-universal formula
\begin{eqnarray}
\label{non-universal-Mfixed}
& & K(z_1,z_2) \nonumber \\ 
& \sim &  \displaystyle  \frac{N^2}{\pi |s_1 s_2|} \left\{ 
\frac{e^{-\gamma (t_1 + t_2)}}{\displaystyle \int_0^T dt 
\ \frac{e^{-2 \gamma t}}{u-t}}
+ s_1 {\bar s}_2 \int_0^1 d c \ 
(c + \gamma) \frac{e^{-(c +\gamma) (t_1 + t_2)} e^{i c (\phi_1 - \phi_2)}}{1 - e^{- 2 (c + \gamma) T}} 
\right\}, \nonumber \\ & & N \rightarrow \infty
\end{eqnarray}
for $e^{i M \psi} = 1$. The breakdown of the universality results from the singularities of the 
potential function (\ref{potential-and-weight})  at the 
points satisfying $z^M = 1$. A subsequent limit $u \rightarrow \infty$ of the non-universal formula (\ref{non-universal-Mfixed}) 
pushes the annulus far away from the unit circle $|z|=1$, and thus recovers the the universal formula (\ref{universal-Mfixed}), as expected.
\par
Let us examine the one-dimensional limit $T \rightarrow 0$ of the non-universal formula 
(\ref{non-universal-Mfixed}) in the case $e^{i M \psi} = 1$. Note that the limits $t_1,t_2 \rightarrow 0$ 
must be taken when we calculate the limit $T \rightarrow 0$. Then we find
\begin{eqnarray}
K_(z_1,z_2) & \sim & \frac{N^2}{\pi T \sqrt{u^2 + \phi_1^2} \sqrt{u^2 + \phi_2^2}} \nonumber \\ 
& & \times \left\{ u + 
\frac{(u + i \phi_1)(u - i \phi_2)}{2} e^{i (\phi_1 - \phi_2)/2} 
\frac{\sin((\phi_1 - \phi_2)/2)}{(\phi_1 - \phi_2)/2} \right\}.  \nonumber \\
\end{eqnarray}

\subsection{Large number of negative charges ($M<N$)}

Let us next consider a large number ($M = O(N)$) of negative charges on the 
unit circle $|z| =1$. In order to first investigate the case with $M < N$, we set
\begin{equation}
\mu = \lim_{N \rightarrow \infty} (M/N), \ \ \ 0 < \mu \leq 1
\end{equation}
and utilize the scaling variables $T$ and $t$ defined in (\ref{scaling-T-t}). As $v > 1$, 
we find
\begin{equation}
r^{- 2 M} = v^{-2 M} \left( 1 - \frac{t}{N} \right)^{- 2 M} \sim 0.
\end{equation}
Then it follows that
\begin{equation}
h_n \sim \frac{\pi v^{2 (n + \Gamma - M)+2}}{N (c + \gamma - \mu)} g(v) 
\left( 1 - e^{-2(c + \gamma - \mu) T} \right), \ \ \ N \rightarrow \infty,
\end{equation}
if $c = \lim_{N \rightarrow \infty} (n/N) $ and $\gamma = \lim_{N \rightarrow \infty} (\Gamma/N)$ 
are fixed. We assume that $g(r)$ is bounded and continuous for $r \leq v$, and suppose that $g(v) > 0$.  
Using the scaling variables $t_1,t_2,\phi_1$ and $\phi_2$ defined in (\ref{scaling}), we obtain
\begin{eqnarray}
\label{mlarge-k1-k2}
K^{(1)}(z_1,z_2) & \sim & 
\frac{N^2}{\pi v^2} \int_0^\mu dc \ (c + \gamma - \mu) \frac{e^{- (c + \gamma - \mu)(t_1 + t_2)} 
e^{i c (\phi_1 - \phi_2)}}{ 1 - e^{- 2 (c + \gamma - \mu) T}}, \nonumber \\ 
K^{(2)}(z_1,z_2) & \sim & 
\frac{N^2}{\pi v^2} \int_\mu^1 dc \ (c + \gamma - \mu) \frac{e^{- (c + \gamma -\mu)(t_1 + t_2)} 
e^{i c (\phi_1 - \phi_2)}}{ 1 - e^{- 2 (c + \gamma -\mu) T}}. 
\end{eqnarray} 
These asymptotic formulas are added to give
\begin{equation}
\label{universal-mlarge}
K(z_1,z_2) \sim 
\frac{N^2}{\pi v^2} \int_0^1 dc \ (c + \gamma-\mu) \frac{e^{- (c + \gamma-\mu)(t_1 + t_2)} 
e^{i c (\phi_1 - \phi_2)}}{ 1 - e^{- 2 (c + \gamma-\mu) T}}.
\end{equation}
This is the universal form (\ref{universal}), when $\gamma$ is replaced with $\gamma - \mu$. The universality still essentially holds, even 
if there is a  large number of negative charges on the unit circle $|z|=1$. 
\par
As before, when the thin annulus is in the outer neighborhood of the unit circle, the universality is broken. Let us again employ the 
scaling variables defined in (\ref{scaling-T-u}). We assume that 
$g(r)$ is bounded and continuous for $r \geq 1$, and suppose that $g(1) > 0$. When $n < M$, 
we can see from (\ref{hn}) that
\begin{equation}
h_n \sim \frac{2 \pi g(1) e^{2 (c + \gamma - \mu) u}}{N} 
\int_0^T dt \frac{e^{- 2 (c + \gamma - \mu) t}}{1 - e^{- 2 \mu (u-t)}}, 
\ \ \ N \rightarrow \infty.
\end{equation} 
When $n \geq M$, we find
\begin{equation}
h_n \sim \frac{\pi g(1) e^{2 (c + \gamma - \mu) u}}{N (c + \gamma - \mu)} 
\left( 1 - e^{-2 (c + \gamma  - \mu) T} \right), \ \ \ N \rightarrow \infty.
\end{equation}
\par
In order to see the outer neighborhood of the unit circle, we employ the scaling variables 
$t_1,t_2,\phi_1$ and $\phi_2$ defined in (\ref{scaling1}). Now $\psi$ in (\ref{scaling1}) 
is supposed to take the values
\begin{equation}
\psi = \frac{2 \pi}{M} k, \ \ \ k=0,1,2,\cdots,M-1,
\end{equation}
which satisfy the non-universality condition $e^{i M \psi } = 1$. Then, if we put
\begin{equation}
\psi \leq \theta_j = \psi + \frac{\phi_j}{N} < \psi + \frac{2 \pi}{M},
\end{equation}
$\theta_j$ is able to take any value on the whole interval $0 \leq \theta_j < 2 \pi$.
Therefore, without loss of generality, we focus on the interval
\begin{equation}
0 \leq \phi_j \leq \frac{2 \pi}{\mu}, \ \ \ j=1,2
\end{equation}
in the asymptotic limit $N \rightarrow \infty$. 
\par
Using $s_j = u - t_j + i \phi_j$ ($j=1,2$), 
we obtain an asymptotic result
\begin{eqnarray}
\label{mlarge-k1}
K^{(1)}(z_1,z_2) & \sim & 
\frac{N^2}{2 \pi |1 - e^{- \mu s_1}| |1 - e^{- \mu s_2}|} \int_0^\mu dc \ \frac{e^{-(c + \gamma - \mu)(t_1 + t_2)} e^{i c (\phi_1 - \phi_2)}}
{\displaystyle \int_0^T dt \ \frac{e^{- 2 (c + \gamma-\mu) t}}{1 - e^{- 2 \mu(u - t)}}}, \nonumber \\ 
& & N \rightarrow \infty.
\end{eqnarray}
Let us take the limit $\mu \rightarrow 0$ of (\ref{mlarge-k1}). It follows from
\begin{equation}
|1 - e^{- \mu s_j}|  \sim \mu |s_j|, \ \ \ \mu \rightarrow 0
\end{equation} 
($j=1,2$) that
\begin{equation}
\label{mlarge-musmall-k1}
K^{(1)}(z_1,z_2) \sim \frac{N^2 e^{- \gamma (t_1 + t_2)}}{\pi |s_1 s_2|} 
\frac{1}{
\displaystyle \int_0^T dt \ \frac{e^{- 2 \gamma t}}{u - t}}, \ \ \ \mu \rightarrow 0.
\end{equation}
This is an expected result, because it is identical to (\ref{non-universal-k1}) for 
$e^{i M \psi} = 1$ with a fixed $M$. We next investigate the limit $u \rightarrow \infty$ 
(with $T$ fixed) of (\ref{mlarge-k1}). In this limit the annulus goes far apart from the 
unit circle, and the universal asymptotic formulas are expected to be recovered. As
\begin{equation}
\int_0^T dt \ \frac{e^{- 2 (c + \gamma-\mu) t}}{1 - e^{- 2 \mu(u - t)}} \sim 
\frac{1}{2 (c + \gamma - \mu)} 
(1 - e^{- 2 (c + \gamma - \mu) T} ), \ \ \ u \rightarrow \infty,
\end{equation}
we find
\begin{equation}
\label{mlarge-ularge-k1}
K^{(1)}(z_1,,z_2) \sim \frac{N^2}{\pi} \int_0^\mu dc \ ( c + \gamma - \mu) 
\frac{e^{-(c + \gamma - \mu) (t_1 + t_2) } e^{i c (\phi_1 - \phi_2)}}{
1 - e^{- 2 ( c + \gamma - \mu) T}}, \ \ \ u \rightarrow \infty.
\end{equation}
This is identical to the first of the asymptotic formulas 
(\ref{mlarge-k1-k2}) with $v=1$, as expected. One is also able to obtain the one-dimensional limit $T \rightarrow 0$ (with 
the limits $t_1,t_2 \rightarrow 0$) of (\ref{mlarge-k1}) as
\begin{equation}
\label{mlarge-tsmall-k1}
K^{(1)}(z_1,z_2) \sim \frac{N^2}{2 \pi T} \frac{1 - e^{-2 \mu u}}{|1 - e^{-\mu( u + i \phi_1)}| 
|1 - e^{- \mu (u + i \phi_2)}|} \frac{e^{i \mu (\phi_1 - \phi_2)} - 1}{i (\phi_1 - \phi_2)}, \ \ \ 
T \rightarrow 0.
\end{equation}
\par
In addition, the asymptotic 
formula for $K^{(2)}(z_1,z_2)$ can be derived as
\begin{eqnarray}
\label{mlarge-k2}
& & K^{(2)}(z_1,z_2) \sim 
\frac{N^2 (1 - e^{-\mu s_1}) (1 - e^{-\mu {\bar s_2}})}
{\pi |1 - e^{-\mu s_1}| | 1 - e^{-\mu s_2}|} \nonumber \\ 
& & \times \int_\mu^1 dc \ ( c + \gamma - \mu) 
\frac{e^{-(c + \gamma - \mu) (t_1 + t_2) } e^{i c (\phi_1 - \phi_2)}}{
1 - e^{- 2 ( c + \gamma - \mu) T}}, \ \ \ N \rightarrow \infty.
\end{eqnarray}
Because of the relation 
\begin{equation}
1 - e^{-\mu s_1} \sim \mu s_1, \ \ \ 
1 - e^{-\mu {\bar s_2}} \sim \mu {\bar s_2}, \ \ \ \mu \rightarrow 0,
\end{equation}
(\ref{mlarge-k2}) in the limit $\mu \rightarrow 0$ is
\begin{equation}
\label{mlarge-musmall-k2}
K^{(2)}(z_1,z_2) \sim 
\frac{N^2 s_1 {\bar s_2}}
{\pi |s_1 s_2|} 
\int_0^1 dc \ ( c + \gamma) 
\frac{e^{-(c + \gamma) (t_1 + t_2) } e^{i c (\phi_1 - \phi_2)}}{1 - e^{- 2 ( c + \gamma) T}}, \ \ \ 
\mu \rightarrow 0.
\end{equation}
As expected, this is identical to (\ref{non-universal-k2}) for $e^{i M \psi} = 1$ with a fixed $M$. 
Let us also derive the limit $u \rightarrow \infty$ of (\ref{mlarge-k2}) as
\begin{equation}
\label{mlarge-ularge-k2}
K^{(2)}(z_1,,z_2) \sim \frac{N^2}{\pi} \int_\mu^1 dc \ ( c + \gamma - \mu) 
\frac{e^{-(c + \gamma - \mu) (t_1 + t_2) } e^{i c (\phi_1 - \phi_2)}}{
1 - e^{- 2 ( c + \gamma - \mu) T}}, \ \ \ u \rightarrow \infty, 
\end{equation}
which is equal to the second of (\ref{mlarge-k1-k2}) with $v = 1$. Moreover the one-dimensional 
limit $T,t_1,t_2 \rightarrow 0$ of (\ref{mlarge-k2}) is
\begin{eqnarray}
\label{mlarge-tsmall-k2}
K^{(2)}(z_1,z_2) & \sim & \frac{N^2}{2 \pi T} 
\frac{
(1 - e^{-\mu (u + i \phi_1)}) 
(1 - e^{-\mu (u - i \phi_2)})}
{| 1 - e^{-\mu (u + i \phi_1)} |
| 1 - e^{-\mu (u + i \phi_2)}|}
\frac{e^{i(\phi_1 - \phi_2)} - e^{i \mu(\phi_1 - \phi_2)}}{
i (\phi_1 - \phi_2)}, \nonumber \\ 
& & T \rightarrow 0.
\end{eqnarray}
\par
The asymptotic formulas  for $K(z_1,z_2)$ are just the sums of 
(\ref{mlarge-k1}) and (\ref{mlarge-k2}), 
(\ref{mlarge-musmall-k1}) and (\ref{mlarge-musmall-k2}),
(\ref{mlarge-ularge-k1}) and (\ref{mlarge-ularge-k2}),
(\ref{mlarge-tsmall-k1}) and (\ref{mlarge-tsmall-k2}),
respectively. In particular, the sum of (\ref{mlarge-ularge-k1}) and (\ref{mlarge-ularge-k2}) gives
\begin{equation}
\label{universal-mlarge1}
K(z_1,,z_2) \sim \frac{N^2}{\pi} \int_0^1 dc \ ( c + \gamma - \mu) 
\frac{e^{-(c + \gamma - \mu) (t_1 + t_2) } e^{i c (\phi_1 - \phi_2)}}{
1 - e^{- 2 ( c + \gamma - \mu) T}}.
\end{equation}
This is equal to the universal formula (\ref{universal-mlarge}) with $v=1$. Moreover, 
we can see from the sum of (\ref{mlarge-tsmall-k1}) and (\ref{mlarge-tsmall-k2}) that the correlation 
functions in the one-dimensional limit do not  depend on $\gamma$ at all.

\subsection{Large number of negative charges ($M \geq N$)}

Next we suppose that the number of negative charges on the unit circle $|z| =1$ 
is larger ($M = O(N)$ and $M \geq N$). We set the parameter $\mu$ as
\begin{equation}
\mu = \lim_{N \rightarrow \infty} (M/N), \ \ \ \mu \geq 1.
\end{equation}
Note that in this case $p_n(z)$ with $n \geq M$ do not appear in the 
formula (\ref{kernel-annulus}) of  the kernel function $K(z_1,z_2)$. Therefore 
we only need the monomial orthogonal polynomials $p_n(z) = z^n$ and
\begin{equation}
h_n = 2 \pi \int_R^v dr \ g(r) \frac{r^{2 (\Gamma - M + n)+1}}{1 - r^{- 2 M}}.
\end{equation}
\par
As before we assume that $g(r)$ is bounded and continuous for $r \leq v$ ($v > 1$) with $g(v) > 0$. 
Consequently, by using the scaling variables in (\ref{scaling}) and (\ref{scaling-T-t}), 
we obtain the universal formula (\ref{universal-mlarge}). 
\par
On the other hand, let us suppose that $g(r)$ is bounded and continuous for $r \geq 1$ with $g(1) > 0$. When one uses the scaling variables in (\ref{scaling-T-u}) and (\ref{scaling1}) satisfying $\psi = 2 \pi k/M$ ($k=0,1,2,\cdots,M-1$), it can readily be seen in the outer 
vicinity of the unit circle that
\begin{equation}
\label{mverylarge}
K(z_1,z_2) \sim \frac{N^2}{2 \pi |1 - e^{-\mu s_1}| |1 - e^{-\mu s_2}|} \int_0^1 dc \ \frac{e^{-(c + \gamma - \mu)(t_1 + t_2)} e^{i c (\phi_1 - \phi_2)}}
{\displaystyle \int_0^T dt \ \frac{e^{- 2 (c + \gamma-\mu) t}}{1 - e^{- 2 \mu(u - t)}}}, \ \ \ N \rightarrow \infty,
\end{equation}
where $0 \leq \phi_j \leq 2 \pi/\mu$ and $s_j = u - t_j + i \phi_j$ ($j=1,2$). 
Both of  the limits $\mu \rightarrow \infty$ (with $\gamma - \mu$ fixed) and the limit $u \rightarrow \infty$ of (\ref{mverylarge}) reproduce the universal formula (\ref{universal-mlarge1}). When we put 
$\mu = 1$, the result is equal to the sum of (\ref{mlarge-k1}) and (\ref{mlarge-k2}), as expected. 
We finally evaluate the one-dimensional limit $T \rightarrow 0$ (with $t_1,t_2 \rightarrow 0$) 
of (\ref{mverylarge}) and find
\begin{equation}
\label{mverylarge-tmall}
K(z_1,z_2) \sim \frac{N^2}{2 \pi T} \frac{(1 - e^{-2 \mu u}) e^{i(\phi_1 - \phi_2)/2} }{|1 - e^{-\mu( u + i \phi_1)}| 
|1 - e^{- \mu (u + i \phi_2)}|} 
\frac{\sin((\phi_1 - \phi_2)/2)}{(\phi_1 - \phi_2)/2}, \ \ \ T \rightarrow 0, 
\end{equation}
which recovers the universal sine kernel formula (\ref{sine-kernel}) with $v = 1$ in the limit $\mu \rightarrow \infty$ or $u \rightarrow \infty$. 
As before the one-dimensional limit (\ref{mverylarge-tmall}) does not depend on $\gamma$ at all.

\section{Annulus in the interior of the unit circle}
\setcounter{equation}{0}
\renewcommand{\theequation}{4.\arabic{equation}}

In \S 3, we assume that the gas molecules are distributed on an annulus in 
the exterior of the unit circle $|z| = 1$. In this section we consider the molecules on an 
annulus in the interior of the unit circle.  Namely, an inner radius $R$ and an 
outer radius $v$ of the annulus satisfy the relation $0< R < v < 1$. We again suppose 
that a point charge $\Gamma$ is at the origin and that a negative unit charge is 
at each of the regular polygon vertices $z = \omega^j$ ($j=0,1,2,\cdots,M-1$) 
with $\omega = e^{2 \pi i/M}$. 
\par
In the case $M < N$, one can show from the orthogonality relation (\ref{B-orthogonality}) 
in Appendix B that the Type B orthogonal polynomials
\begin{equation}
\label{OP-in}
p_n(z) = \left\{ \begin{array}{ll} z^n (1 - z^M), & 0 \leq n < N-M, 
\\ z^n, & N-M \leq n < N \end{array} \right.
\end{equation}
satisfy the orthogonality relation
\begin{equation}
\int_0^{2 \pi} d\theta \ |D(z)|^2 \ p_m({\bar z})  p_n(z) = h_n \delta_{mn}, \ \ \ m,n=0,1,2,\cdots,N-1,
\end{equation}
where $z = r e^{i \theta}$ ($0 < r = |z| < 1$), $D(z) = 1/(1 - z^{-M})$ and
\begin{equation}
h_n = \left\{ \begin{array}{ll} 2 \pi r^{2 n + 2 M}, & 0 \leq n < N-M,  \\ 
\displaystyle \frac{2 \pi r^{2 n}}{r^{-2 M} - 1}, & N-M \leq n < N.  \end{array} \right.
\end{equation}
Then the kernel function defined in (\ref{kernel}) on the annulus has the form
\begin{equation}
K(z_1,z_2) = \sqrt{g(r_1)  g(r_2)} (r_1 r_2)^{\Gamma-M} |D(z_1) D(z_2)| \sum_{n=0}^{N-1} \frac{1}{h_n} p_n(z_1) p_n( {\bar z_2})
\end{equation}
with $0< R \leq r_1,r_2 \leq v < 1$ ($r_1=|z_1|$, $r_2 = |z_2|$). Here $p_ n(z)$ defined in (\ref{OP-in}) satisfy
\begin{eqnarray}
& & \int d^2z \ g(r) \ r^{2 (\Gamma - M)} \  |D(z)|^2 p_m({\bar z}) p_n(z)  \nonumber \\
& = & \int_R^v dr  \int_0^{2 \pi} d\theta \ r^{2(\Gamma - M)+1} \ g(r) \ |D(z)|^2 \ p_m(\bar{z})  p_n(z) = h_n \delta_{mn}
\end{eqnarray}
with
\begin{equation}
h_n =  \left\{ \begin{array}{ll} \displaystyle
2 \pi \int_R^v dr \ g(r) r^{2 (\Gamma + n) + 1}, & 0 \leq n < N-M, \\
\displaystyle 
2 \pi \int_R^v dr \ g(r) \frac{r^{2 (\Gamma - M + n) + 1}}{r^{-2 M} - 1}, 
& N-M \leq n < N. \end{array} \right.
\end{equation}
Then $K(z_1,z_2)$ is separated into two parts as
\begin{equation}
K(z_1,z_2) = K^{(1)}(z_1,z_2) + K^{(2)}(z_1,z_2),
\end{equation}
where
\begin{equation}
K^{(1)}(z_1,z_2) = \frac{\sqrt{g(r_1)  g(r_2)} (r_1 r_2)^\Gamma}{|z_1^M - 1| | z_2^M - 1|} 
\sum_{n=0}^{N-M-1} \frac{1}{h_n} 
z_1^n (1 - z_1^M) {\bar z_2}^n (1 - {\bar z_2}^M) 
\end{equation}
and
\begin{equation}
K^{(2)}(z_1,z_2) = \frac{\sqrt{g(r_1)  g(r_2)} (r_1 r_2)^\Gamma}{|z_1^M - 1| | z_2^M - 1|} 
\sum_{n=N-M}^{N-1} \frac{(z_1 {\bar z_2})^n}{h_n}.
\end{equation}

\subsection{Fixed number of negative charges}

We first consider the case $M$ fixed and consider the thin annulus limit $N \rightarrow \infty$ of the 
kernel function. The scaling parameters $T$ and $t$ are as before defined in (\ref{scaling-T-t}). Note that this 
time we set $v$ fixed with $0 < v < 1$, because the annulus is in the interior of the unit circle $|z|=1$. 
We assume that $g(r)$ is bounded and continuous for $r \leq v$ with $g(v) > 0$.  Then we can see in the limit 
$N \rightarrow \infty$ that
\begin{equation}
h_n \sim 
\left\{ \begin{array}{ll} \displaystyle 
\frac{\pi v^{2(\Gamma + n) + 2}}{N(c + \gamma)}  \left( 1 - e^{- 2(c + \gamma)T} \right) g(v), 
& 0 \leq n < N-M, \\ \displaystyle
\frac{\pi v^{2(\Gamma + n) + 2}}{N(1 + \gamma)}   \left( 1 - e^{- 2(1 + \gamma)T} \right) 
\frac{g(v)}{1 - v^{2 M}}, 
& N-M \leq n < N \end{array} \right.
\end{equation}
with $c = \lim_{N \rightarrow \infty} (n/N)$ and 
$\gamma = \lim_{N \rightarrow \infty} (\Gamma/N)$ fixed. Let us introduce the polar coordinates as $z_1 = r_1 e^{i \theta_1}$, $z_2 = r_2 e^{i \theta_2}$ and use the scaling variables $t_1,t_2,\phi_1$ and $\phi_2$ defined in (\ref{scaling}). In the limit $N \rightarrow \infty$, we find that $K^{(2)}(z_1,z_2)$ is negligible compared to $K^{(1)}(z_1,z_2)$ and consequently find the universal form
\begin{equation}
K(z_1,z_2) \sim \frac{N^2}{\pi v^2} \int_0^1 dc \ ( c + \gamma)  
\frac{e^{-(c + \gamma) (t_1 + t_2)} e^{i c (\phi_1 - \phi_2)}}{
1 - e^{- 2 (c + \gamma) T}},
\end{equation}
which is identical to (\ref{universal}). As is anticipated from the duality relation (\ref{duality-relation}), this universal form is 
invariant  under the mapping
\begin{equation}
\gamma \mapsto - \gamma -1, \ \ \ t_j \mapsto T - t_j, \ \ \ \phi_j \mapsto -\phi_j \ \  \ j=1,2
\end{equation}
aside from a phase factor $e^{-i (\phi_1 - \phi_2)}$, which does not affect the correlation functions.
\par
In order to analyze a thin annulus in the inner neighborhood of the unit circle $|z|=1$, we make use of the scaling 
parameters $u$ and $T$ in (\ref{scaling-T-u}) with $u < 0 < T$. Now $g(r)$ is supposed to be bounded and 
continuous for $r \leq 1$ with $g(1) > 0$.  Then we find an asymptotic relation
\begin{equation}
h_n \sim \left\{ \begin{array}{ll} \displaystyle \frac{\pi g(1) e^{2(c + \gamma) u}}{N (c + \gamma)}  
(1 - e^{-2(c + \gamma) T}), &  0 \leq n < N-M, \\ 
\displaystyle \frac{\pi g(1) e^{2 (1 + \gamma) u}}{M} \int_0^T dt \frac{e^{-2 (1 + \gamma) t}}{t - u}, 
& N-M \leq n < N. \end{array} \right.
\end{equation}
Using the scaling variables $t_1,t_2,\phi_1$ and $\phi_2$ in (\ref{scaling1}),  we find the universal formula 
((\ref{universal}) with $v = 1$)
\begin{equation}
\label{universal-Mfixed-in}
K(z_1,z_2) \sim \frac{N^2}{\pi}  \int_0^1 dc \ (c +\gamma)
\frac{e^{-(c +\gamma) (t_1 + t_2)} e^{i c (\phi_1 - \phi_2)}}{1 - e^{- 2 (c + \gamma) T}}, 
\ \ \ N \rightarrow \infty
\end{equation}
for $e^{i M \psi} \neq 1$, and a non-universal formula
\begin{eqnarray}
\label{non-universal-Mfixed-in}
& & K(z_1,z_2) \nonumber \\ 
& \sim &  \displaystyle  \frac{N^2}{\pi |s_1 s_2|} \left\{ 
\frac{e^{-(1 + \gamma) (t_1 + t_2)} e^{i (\phi_1 - \phi_2)} }{\displaystyle \int_0^T dt 
\ \frac{e^{-2 (1 + \gamma) t}}{t-u}}
+ s_1 {\bar s}_2 \int_0^1 d c \ 
(c + \gamma) \frac{e^{-(c +\gamma) (t_1 + t_2)} e^{i c (\phi_1 - \phi_2)}}{1 - e^{- 2 (c + \gamma) T}} 
\right\}, \nonumber \\ & & N \rightarrow \infty
\end{eqnarray}
for $e^{i M \psi} = 1$. Here $s_j = u - t_j + i \phi_j$ ($j=1,2$). The breakdown of the universality 
again takes place in the neighborhood of the  points satisfying $z^M = 1$. 
\par
In order to make sure of 
the duality relation (\ref{duality-relation}), we apply the mapping
\begin{equation}
\gamma \mapsto - \gamma - 1, \ \ \ t_j \mapsto T  - t_j, \ \ \ \phi_j \mapsto -\phi_j, \ \ \ j=1,2
\end{equation}
and
\begin{equation}
u \mapsto T - u.
\end{equation}
Then the universal formula (\ref{universal-Mfixed-in}) is mapped to (\ref{universal-Mfixed}) and 
the non-universal formula (\ref{non-universal-Mfixed-in}) is mapped to (\ref{non-universal-Mfixed}) 
aside from a phase factor $e^{-i (\phi_1 - \phi_2)}$, as expected.

\subsection{Large number of negative charges ($M<N$)}

Next we study the case with a large number of negative charges. We suppose 
that $M < N$ still holds and there exists
\begin{equation}
\mu = \lim_{N \rightarrow \infty} (M/N), \ \ \ 0 < \mu \leq 1.
\end{equation}
Using the scaling variable $T$ defined in (\ref{scaling-T-t}) and noting $0<v<1$, we can derive
\begin{equation}
h_n \sim \frac{\pi v^{2 (n + \Gamma)+2}}{N (c + \gamma)} g(v) 
\left( 1 - e^{-2(c + \gamma) T} \right), \ \ \ 0 \leq n < N
\end{equation}
in the limit $N \rightarrow \infty$ with $c = \lim_{N \rightarrow \infty} (n/N) $ and $\gamma = 
\lim_{N \rightarrow \infty} (\Gamma/N)$ fixed.  Let us assume that $g(r)$ is bounded and continuous 
for $r \leq v$, and suppose that $g(v) > 0$.  By means of the the scaling variables 
$t_1,t_2,\phi_1$ and $\phi_2$ defined in (\ref{scaling}), one finds
\begin{equation}
K(z_1,z_2) \sim 
\frac{N^2}{\pi v^2} \int_0^1 dc \ (c + \gamma) \frac{e^{- (c + \gamma)(t_1 + t_2)} 
e^{i c (\phi_1 - \phi_2)}}{ 1 - e^{- 2 (c + \gamma) T}}.
\end{equation}
This is the universal form (\ref{universal}), which does not depend on the parameter $\mu$, as a reflection 
of the fact that the negative charges on the unit circle $|z|=1$ are all (sufficiently far) out of the outer boundary 
of the annulus.
\par
The universality is broken in the inner neighborhood of the unit circle. In order to observe 
the breakdown, we utilize the scaling variables defined in (\ref{scaling-T-u}) with $u < 0 < T$. 
We moreover assume that $g(r)$ is bounded and continuous for $r \leq 1$, and suppose that $g(1) > 0$. 
Then in the limit $N \rightarrow \infty$ it follows that
\begin{equation}
h_n \sim \left\{ \begin{array}{ll} \displaystyle 
\frac{\pi g(1)e^{2 (c + \gamma) u} }{N (c + \gamma)} 
\left( 1 - e^{-2 (c + \gamma) T} \right), & 0 \leq n < N-M, \\ 
\displaystyle \frac{2 \pi g(1) e^{2 (c + \gamma) u}}{N} \int_0^T dt \frac{e^{-2 (c + \gamma) t}}{1 
- e^{2 \mu (u-t)}}, 
& N-M \leq n < N. \end{array} \right.
\end{equation} 
\par
Let us employ the scaling variables $t_1,t_2,\phi_1$ and $\phi_2$ defined in (\ref{scaling1}).
As argued in \S 3, without loss of generality we can focus on the interval 
\begin{equation}
0 \leq \phi_j \leq \frac{2 \pi}{\mu}, \ \ \ j=1,2
\end{equation}
with $e^{i M \psi } = 1$. Then, using $s_j = u - t_j + i \phi_j$ ($j=1,2$), we find
\begin{equation}
\label{mlarge-k1-in}
K^{(1)}(z_1,z_2) \sim 
\frac{N^2 (1 - e^{\mu s_1}) (1 - e^{\mu {\bar s_2}})}
{\pi |1 - e^{\mu s_1}| | 1 - e^{\mu s_2}|} \int_0^{1-\mu} dc \ ( c + \gamma) 
\frac{e^{-(c + \gamma) (t_1 + t_2) } e^{i c (\phi_1 - \phi_2)}}{
1 - e^{- 2 ( c + \gamma) T}}
\end{equation}
and
\begin{equation}
\label{mlarge-k2-in}
K^{(2)}(z_1,z_2) \sim 
\frac{N^2}{2 \pi |1 - e^{\mu s_1}| |1 - e^{\mu s_2}|} \int_{1 - \mu}^1 dc \ \frac{e^{-(c + \gamma)(t_1 + t_2)} e^{i c (\phi_1 - \phi_2)}}
{\displaystyle \int_0^T dt \ \frac{e^{- 2 (c + \gamma) t}}{1 - e^{2 \mu(u - t)}}}
\end{equation}
in the limit $N \rightarrow \infty$.
\par
The asymptotic formula for $K(z_1,z_2)$ is the sum of (\ref{mlarge-k1-in}) and (\ref{mlarge-k2-in}). 
The duality relation (\ref{duality-relation}) can be confirmed by introducing the mapping
\begin{equation}
\label{map-in1}
\gamma \mapsto - \gamma + \mu - 1, \ \ \ t_j \mapsto T  - t_j, \ \ \ \phi_j \mapsto -\phi_j, \ \ \ j=1,2
\end{equation}
and
\begin{equation}
\label{map-in2}
u \mapsto T - u.
\end{equation}
We find that $K^{(1)}(z_1,z_2)$ in (\ref{mlarge-k1-in}) is mapped to $K^{(2)}(z_1,z_2)$ in (\ref{mlarge-k2}) and 
$K^{(2)}(z_1,z_2)$ in (\ref{mlarge-k2-in}) is mapped to $K^{(1)}(z_1,z_2)$ in (\ref{mlarge-k1}) aside from 
a phase factor $e^{-i (\phi_1 - \phi_2)}$, as anticipated.

\subsection{Large number of negative charges ($M \geq N$)}

We finally examine the case $M \geq N$ with the parameter
\begin{equation}
\mu = \lim_{N \rightarrow \infty} (M/N), \ \ \ \mu \geq 1.
\end{equation}
In this case the corresponding orthogonal polynomials $p_n(z)$ ($0 \leq n < N$) are monomials 
$z^n$ (see (\ref{MgeqN})) and
\begin{equation}
h_n = 2 \pi \int_R^v dr \ g(r) \frac{r^{2 (\Gamma - M + n)+1}}{r^{- 2 M} - 1}, \ \ \ 0 \leq n < N.
\end{equation}
Assuming that $g(r)$ is bounded and continuous for $r \leq v$ ($0 < v  < 1$) with $g(v) > 0$ and 
using the scaling variables in (\ref{scaling-T-t}), we can derive 
\begin{equation}
h_n \sim \frac{\pi v^{2 (n + \Gamma) + 2}}{N (c + \gamma)} g(v) \left( 1 - e^{-2( c + \gamma ) T} \right)
\end{equation}
in the limit $N \rightarrow \infty$ with $c = \lim_{N \rightarrow \infty} (n/N) $ and $\gamma = 
\lim_{N \rightarrow \infty} (\Gamma/N)$ fixed. Then, as for the the kernel function (\ref{kernel-annulus}), 
one utilizes the scaling variables in (\ref{scaling}) and obtains the universal formula (\ref{universal}) in the thin annulus limit.
\par
Let us next consider a thin annulus in the inner vicinity of the unit circle $|z|=1$. For that purpose, we 
use the scaling variables in (\ref{scaling-T-u}) with $u < 0 < T$ and (\ref{scaling1}) with $e^{i M \psi} = 1$. 
Supposing that $g(r)$ is bounded and continuous for $r \leq  1$ with $g(1) > 0$, one can readily find
\begin{equation}
h_n \sim \frac{2 \pi g(1)}{N} e^{2 (c + \gamma - \mu) u} \int_0^T dt \frac{e^{-2 (c + \gamma - \mu)t}}{e^{-2 \mu (u - t)} - 1}
\end{equation}
and
\begin{equation}
\label{mverylarge-in}
K(z_1,z_2) \sim \frac{N^2}{2 \pi |1 - e^{-\mu s_1}| |1 - e^{-\mu s_2}|} \int_0^1 dc \ \frac{e^{-(c + \gamma - \mu)(t_1 + t_2)} e^{i c (\phi_1 - \phi_2)}}
{\displaystyle \int_0^T dt \ \frac{e^{- 2 (c + \gamma-\mu) t}}{e^{- 2 \mu(u - t)}-1}}, \ \ \ N \rightarrow \infty.
\end{equation}
Here $0 \leq \phi_j \leq 2 \pi/\mu$ and $s_j = u - t_j + i \phi_j$ ($j=1,2$). The mappings (\ref{map-in1}) and (\ref{map-in2}) map this asymptotic form (\ref{mverylarge-in}) to (\ref{mverylarge}) aside from a phase factor $e^{-i (\phi_1 - \phi_2)}$, as is deduced from the duality relation (\ref{duality-relation}).

\section{Summary and discussion}
\setcounter{equation}{0}
\renewcommand{\theequation}{5.\arabic{equation}}

In this paper, two-dimensional one-component Coulomb gases were analyzed, and 
the asymptotic correlation functions among the gas molecules in the 
thermodynamic limit were evaluated at a special inverse temperature $\beta=2$. 
The gas molecules were distributed on an annulus around the origin and each of them 
was supposed to carry a unit positive charge. In addition, we put point charges at 
some fixed locations. We first put one point charge at the origin and derived 
universal forms of the correlation functions. We next added negative point 
charges on the unit circle around the origin and derived non-universal forms 
of the correlation functions, when the annulus was contained in the neighborhood 
of the unit circle. We obtained a duality relation of the correlation functions in 
Appendix A and found it useful to treat an annulus in the interior of the unit circle.
The duality relation is valid for general $\beta > 0$, and it is interesting to make clear 
the implications of the duality for general Coulomb gas systems.
\par
An annulus (\ref{annulus}) has an inner hard wall (at $|z|=R$) and an outer hard wall (at $|z|=v$). 
As mentioned in \S 2, an outer hard wall (hard edge) naturally appears in the theory of truncated random unitary matrices\cite{ZS}. Let us consider an $N \times N$ sub-block of an $N' \times N'$ random unitary matrix with $N'>N$. The $N$ eigenvalues of the sub-block are distributed on the disc 
$|z| \leq 1$ with a probability density function (\ref{pdf}) with $\beta=2$ and
\begin{equation}
\label{truncated}
w(z) = (1 - |z|^2)^{\nu-1}, \ \ \ \nu=N'-N.
\end{equation}
This is in the class (I) and the corresponding kernel function has the form (\ref{classI}).  
Note that a two-dimensional Coulomb gas on a  flat disc is recovered in a special case $\nu=1$. 
In particular the eigenvalue density is given by
\begin{equation}
\rho(z) = K(z,z) = \frac{1}{\pi} (1 - |z|^2)^{\nu-1} \sum_{n=0}^{N-1} \frac{(\nu+n)!}{n! (\nu-1)!} |z|^{2 n}.
\end{equation}
Then, in the limit $N \rightarrow \infty$ with $\nu$ as well as $|z| < 1$ fixed, 
we obtain the bulk density \cite{ZS,FS}
\begin{equation}
\label{bulk}
\rho(z) \sim \frac{\nu}{\pi} \frac{1}{(1 -|z|^2)^2}.
\end{equation}
This is in agreement with (\ref{dual-kI-as}), when $\nu=1$, $\Gamma=0$ and $v=1$. As before, 
in order to magnify the edge region, we introduce a scaling variable $t$ as
\begin{equation}
|z| = 1 - \frac{t}{N}.
\end{equation}
Then we find an asymptotic form\cite{ZS,FS}
\begin{equation}
\rho(z) \sim \kappa_{\rm hard}(z) = \frac{N^2}{\pi (\nu-1)!} (2 t)^{\nu-1} \int_0^1 dc \ c^\nu e^{-2 c t}
\end{equation}
with $\nu$ fixed. This is again in agreement with (\ref{kappa}), when $\nu=1$, $\gamma=0$ and $v=1$. In \S 2,  we 
show in (\ref{kappa1}) and (\ref{kappa3}) that this asymptotic form with $\nu=1$ is also observed 
at the outer edge of an annulus, and it is essentially observed at the inner edge as well, if one takes 
into account the duality explained in Appendix A. As the asymptotic form satisfies
\begin{equation}
\int \kappa_{\rm hard}(z) d^2z = \frac{2 \pi}{N} \int_0^\infty dt \ \frac{N^2}{\pi (\nu-1)!} (2 t)^{\nu-1} \int_0^1 dc \ c^\nu e^{-2 c t} = N,
\end{equation}
one can see that almost all the eigenvalues are accumulated in the narrow edge region 
$1 - |z| = O(1/N)$. In other words, the contribution from the bulk region is negligibly small. 
This hollow structure is a typical feature of a two-dimensional Coulomb gas confined by 
a hard wall. 
\par
On the other hand, in Ginibre's work\cite{ginibre} on non-hermitian random matrices, 
a model with a substantial contribution from the bulk density was treated. It is also in  the 
class (I) with $|z| < \infty$ and a weight function
\begin{equation}
w(z) = e^{- |z|^2},
\end{equation}
which confines molecules around the origin. It can be considered as the limit $\nu \rightarrow \infty$ of (\ref{truncated}) after putting $|z| \rightarrow |z|/\sqrt{\nu}$. The corresponding eigenvalue density is
\begin{equation}
\rho(z) = K(z,z) = \frac{1}{\pi} e^{- |z|^2} \sum_{n=0}^{N-1} \frac{(|z|^2)^n}{n!}
\end{equation}
and the limit $N \rightarrow \infty$ is known to be\cite{ginibre,mehta}
\begin{equation}
\label{ginibre-bulk}
\rho(z) \sim \kappa_{\rm bulk}(z) = \left\{ \begin{array}{ll} 1/\pi,  & |z| < \sqrt{N}, \\ 
0, &  |z| > \sqrt{N}. \end{array} \right.
\end{equation}
Since this asymptotic form satisfies
\begin{equation}
\int \kappa_{\rm bulk}(z) d^2z = 2 \pi \int_0^{\sqrt{N}} \frac{r}{\pi} \ dr = N,
\end{equation}
we can say that almost all the eigenvalues are found in the bulk region. 
Such a solid bulk region with a substantial contribution to the total number 
of eigenvalues is called a droplet. 
\par
At the edge $|z| = \sqrt{N}$ of the droplet of Ginibre's model, the eigenvalues are 
not strictly confined within a hard wall, so that a small number of them penetrate 
into the outer region. We call such an edge a soft edge. In order to magnify the 
soft edge region of Ginibre's model, one employs the scaling variable $s$ as
\begin{equation}
|z| = \sqrt{N} + s.
\end{equation}
In the limit $N \rightarrow \infty$ one finds\cite{jancovici1,jancovici2}
\begin{equation}
\label{kappa-soft}
\rho(z) \sim \kappa_{\rm soft}(z) = \frac{1}{2 \pi} {\rm erfc}(\sqrt{2} s),
\end{equation}
where ${\rm erfc}(x)$ is the complementary error function. It follows that 
the number of the eigenvalues in the soft edge region 
$\sqrt{N} - |z| = O(1)$  is $O(\sqrt{N})$, which 
is negligibly small compared to the total number of the eigenvalues. The asymptotic 
formula (\ref{kappa-soft}) is now known to be universal\cite{HW} and characterizes 
a feature of general soft edges. The asymptotic behavior of a solid droplet with 
soft edges is thus greatly different from that of a hollow Coulomb gas confined by 
hard walls in both bulk and edge regions\cite{CW,DS}. 
\par
Instead of (or in addition to) soft edges, one can also equip droplets with hard walls.  
A typical example is the hard wall Ginibre model\cite{jancovici1,jancovici2} which 
has a restriction $|z| \leq \sqrt{N}$ in addition to a confinement due to 
a weight function $w(z) = e^{-|z|^2}$. In this model, the hard wall 
is put at the edge of a droplet. As a result, the asymptotic form of the bulk density 
keeps the form (\ref{ginibre-bulk}), although the edge behavior (\ref{kappa-soft}) is modified. Thus one obtains a solid droplet equipped with a hard wall. Such models have extensively been studied in recent works\cite{CW,ameur,AKM,AKMW,AKS1,seo2020,seo2022,AFLS}. Moreover droplets can be divided into several domains or split into several disconnected parts. Such a model with discontinuities or gaps is another major subject of current 
research\cite{charlier2, ACC2,noda}.
\par
In this paper, we focus on two-dimensional Coulomb gases confined by hard walls, 
putting several point charges in the region out of the walls. Let us explain physical 
motivations to consider such situations.
\par
\medskip
\noindent
(1) Action at a distance
\par
\medskip
\noindent
There are many works on point charges put in a Coulomb gas\cite{LY1,WW,LY2,AKS2,byun1,byun2,BKSY,KKL,DMMS,BFL,BFKL}. 
The effect of such point charges on the distribution of nearby gas molecules is 
physically interesting and important. It depends on a subtle balance of the force 
originating from the point charges and the Coulomb interaction among gas molecules. 
In this paper, point charges are put in the region out of the hard walls, so that 
they are strictly separated from the Coulomb gas.  Such point charges can still 
affect the gas, because the classical Coulomb interaction is an action at a distance.
However, when the sum of the point charges is fixed, it is negligibly small compared 
to the total macroscopic charge of the gas. In that situation, the point charges hardly 
modify the statistical behavior of the gas molecules. 
\par
In order to strengthen the action at a distance on a remote gas, 
one can introduce a macroscopic charge\cite{fischmann,BFL,BFKL} 
proportional to the number $N$ of the gas molecules. Suppose that 
there is a positively charged gas on an annulus centered at the origin. 
If we put a positive or zero point charge at the origin, the molecules are 
accumulated at the outer edge of the annulus due to 
the Coulomb repulsion (see (\ref{kappa1})). When a macroscopic negative charge 
with a sufficiently large absolute value is put at the origin, the molecules 
are drawn to the origin and accumulated at the inner edge (see (\ref{kappa3})). 
Moreover, when a macroscopic negative charge with a small absolute value is 
put at the origin, the repulsion among the gas molecules and the attraction at a distance 
from the origin are balanced, so that  the gas molecules are accumulated at both outer 
and inner edges of the annulus (see (\ref{kappa2}) and (\ref{kappa2tilde})). 
One of the major physical motivations to set up hard walls and point charges out of 
the walls is the observation of such a qualitative outcome of the action at a distance.
\par
\medskip
\noindent
(2) Universality breakdown
\par
\medskip
\noindent
The second physical motivation comes from the breakdown of universality. The universal 
form of  the molecule correlation functions on a thin annulus is given in (\ref{universal}). 
We attempt to break the universality by means of an action at a distance from external 
charges. How should the external charges be placed in the region out of the hard wall?  
\par
We first consider a charge distribution with a continuous rotational symmetry. Let us suppose 
that an external charge is uniformly distributed on a circle centered at the origin with a radius $r_0 > 0$. If the angular density of the external charge is $q$, the Coulomb potential acting on a unit charge of a gas molecule at the location $r > 0$ ($r \neq r_0$) is
\begin{equation}
\Phi(r) = -q \int_0^{2 \pi} d\theta \ \log| z - r |, \ \ \ z = r_0 e^{i \theta}.
\end{equation}
Then it can be seen that
\begin{equation}
\frac{d}{d r} \Phi(r) =  \frac{q}{r}  \ {\rm Im} \oint dz \left( \frac{1}{z - r} - \frac{1}{z} \right).
\end{equation}
Here the last integral is taken over the contour anticlockwisely going on the circle centered at the 
origin with a radius $r_0$. Because of Cauchy's residue theorem, we obtain
\begin{equation}
\frac{d}{d r} \Phi(r) = \left\{ \begin{array}{ll} \displaystyle 
- \frac{Q}{r}, & {\rm if} \ r > r_0, \\ 
0, & {\rm if} \ r < r_0, \end{array} \right.
\end{equation}
where $Q = 2 \pi q$ is the total external charge. Hence the gas molecule at the location $r$ 
feels as if a point external charge $Q$ was put at the origin, if $r > r_0$.  On the other hand, if  $r < r_0$, 
it feels as if there were no external charges.
\par
This result implies the following. Suppose that external charges are distributed in the 
region out of the annulus. If the charge distribution is invariant under a continuous rotation around 
the origin, it can be replaced with a point charge at the origin without essentially 
changing the Coulomb potential on the annulus. Accordingly, a charge distribution with a continuous 
rotational symmetry keeps the universal form (\ref{universal}), except a change of the parameter 
value $\gamma$. Therefore such a charge distribution does not break the universality. 
\par
Another possible way to break the universality is a smooth deformation of the hard wall 
boundary. However, from the analysis of the class (III), (IV) and (V) Coulomb gases, the 
correlation functions of the gas molecules on an asymmetrically deformed elliptic annulus 
still keep the universal form\cite{elliptic}, at least in the case $\gamma=0$. Moreover, 
in the limit $T \rightarrow \infty$, one recovers the universal form (\ref{kappa}) with 
$\gamma=0$,  which is known to hold in the neighborhood of a general smooth curve of 
a hard wall\cite{CW}. It therefore seems that a discontinuity or singularity of the potential function 
should be introduced in order to observe a universality breakdown. 
\par
A charge distribution with a discrete rotational symmetry gives one of the simplest models with 
such a singularity. A model in  the class (II) is an example including negative 
point charges producing a potential function with singularities. A detailed study in \S 3 
and \S 4 results in non-universal forms of correlation functions such as  
(\ref{non-universal-Mfixed}). These give explicit quantitative evidences of a universality 
breakdown. However, such 
a non-universal form holds only in the neighborhood of the potential singularities, where 
the non-universality condition $e^{i M \psi}=1$ is satisfied. In an attempt to 
strengthen the breakdown, we also consider a macroscopic number of negative charges 
($M=O(N)$). But the breakdown takes place only in the neighborhood of the (infinitely many 
but discrete) singularities as before. This result demonstrates that the universality is robust and can be broken only under restrictive conditions. 
\par
Several Coulomb gas models with discrete rotational symmetries have also been studied in the case of droplets\cite{BM,BGM,BY,byun3}. In particular, the soft-edge universal form (\ref{kappa-soft}) was derived in the asymptotic analysis of the corresponding kernel functions, 
even if there were several disconnected droplets\cite{BY}. This also provides evidence for the robustness of universality.
\par
\medskip
\noindent
(3) One-dimensional limit
\par
\medskip
\noindent
The third physical motivation is the observation of  the transition between two-dimensional 
and one-dimensional systems.  The circular unitary ensemble (CUE) is a renowned 
model of random unitary matrices\cite{dyson}. The CUE eigenvalues are on a unit circle and 
interact with each other via two-dimensional Coulomb force. Such a gas should 
be realized on a circle embedded into a two-dimensional substrate. 
In that situation, a quasi one-dimensional ring with a small but non-zero width 
is more realistic than a strictly one-dimensional circle. This motivates the analysis 
of a thin annulus sandwiched between hard walls. In the one-dimensional limit 
$T \rightarrow 0$, we can recover the CUE sine-kernel formula (\ref{sine-kernel}).
\par
On the other hand, a droplet ring with a non-zero width and a uniform density is known to be formed in the induced ensemble of random matrices\cite{FBKSZ}. The one-dimensional (small width) limit of 
such droplets has been studied in \cite{BS}. A hollow annulus between hard walls analyzed in this 
paper and such a solid droplet have a great difference in molecule density profiles. Nevertheless both 
of them finally recover the same sine-kernel formula in the one-dimensional limit.

\section*{Acknowledgements}

The author acknowledges support by the Japan Society for the Promotion of Science (KAKENHI 20K03764). 
He also thanks Aron Wennman for pointing to the reference \cite{walsh2}.

\section*{Appendix A} 
\setcounter{equation}{0}
\renewcommand{\theequation}{A.\arabic{equation}}

In this Appendix, we study a duality relation between the Coulomb gases on the annuli 
$\{ z | \ R \leq |z| \leq v \}$ and $\{ z| \ 1/v \leq |z| \leq 1/R \}$ with an inverse 
temperature $\beta > 0$. Let us use a notation
\begin{equation}
{\tilde z}_j = \frac{1}{z_j}
\end{equation}
and begin with 
\begin{equation}
|z_j - z_{\ell}|^{\beta} = \frac{
|{\tilde z}_j - {\tilde z}_{\ell}|^{\beta}
}{|{\tilde z}_j {\tilde z}_{\ell}|^{\beta}}
\end{equation}
for $\beta > 0$. Using the identity
\begin{equation}
\prod_{j=1}^N \prod_{\ell=j+1}^N |{\tilde z}_j {\tilde z}_{\ell} |^{\beta} = \prod_{j=1}^N | {\tilde z}_j |^{\beta (N-1)},
\end{equation}
we obtain
\begin{equation}
\label{tildez}
\prod_{j=1}^N \prod_{\ell=j+1}^N |z_j - z_{\ell}|^{\beta} 
= \left( \prod_{j=1}^N \frac{1}{|{\tilde z}_j|^{\beta (N-1)}} \right) 
\left( \prod_{j=1}^N \prod_{\ell=j+1}^N |{\tilde z}_j - {\tilde z}_{\ell}|^{\beta} \right).
\end{equation}
Moreover it is necessary to compute the Jacobian for the variable transformation 
\begin{equation}
(x,y) \mapsto ({\tilde x},{\tilde y})
\end{equation}
with real $x$,$y$,${\tilde x}$ and ${\tilde y}$, where complex variables $z = x + i y$ and 
${\tilde z} = {\tilde x} + i {\tilde y}$ satisfy ${\tilde z} = 1/z$. It follows from
\begin{equation}
x + i y = \frac{1}{{\tilde x} + i {\tilde y}} = \frac{{\tilde x}}{{\tilde x}^2 + {\tilde y}^2} 
- i \frac{{\tilde y}}{{\tilde x}^2 + {\tilde y}^2}
\end{equation}
that the Jacobian is
\begin{equation}
\label{jacobian}
\left| \frac{\partial(x,y)}{\partial({\tilde x},{\tilde y})} \right|= 
\left| 
\begin{array}{cc} 
\displaystyle 
\frac{\partial x}{\partial {\tilde x}} & 
\displaystyle
\frac{\partial x}{\partial {\tilde y}} \\ 
\displaystyle 
\frac{\partial y}{\partial {\tilde x}} & 
\displaystyle 
\frac{\partial y}{\partial {\tilde y}} 
\end{array} \right| = |z|^4 = \frac{1}{|{\tilde z}|^4}.
\end{equation}
\par
Corresponding to (\ref{pdf}) and (\ref{correlation}), we now introduce 
the probability density function 
\begin{equation}
{\tilde P}({\tilde z}_1,{\tilde z}_2,\cdots,{\tilde z}_N) = \frac{1}{{\tilde Z}_N} 
\prod_{j=1}^N {\tilde w}({\tilde z}_j)
\prod_{j<l}^N |{\tilde z}_j - {\tilde z}_{\ell}|^{\beta}
\end{equation} 
with
\begin{equation}
{\tilde w}({\tilde z}_j) = \frac{w(1/{\tilde z}_j)}{|{\tilde z}_j|^{\beta (N-1) + 4}},
\end{equation}
\begin{equation}
{\tilde Z}_N = 
\int d^2{\tilde z}_1 \int d^2{\tilde z}_2 \cdots \int d^2{\tilde z}_N 
\prod_{j=1}^N {\tilde w}({\tilde z}_j)
\prod_{j<l}^N |{\tilde z}_j - {\tilde z}_{\ell}|^{\beta},
\end{equation}
and the correlation functions
\begin{equation}
\label{dual-correlation}
{\tilde \rho}({\tilde z}_1,{\tilde z}_2,\cdots,{\tilde z}_k) = 
\frac{N!}{(N-k)!}  \int d^2{\tilde z}_{k+1} \int d^2{\tilde z}_{k+2} \cdots \int d^2{\tilde z}_N  
P({\tilde z}_1,{\tilde z}_2,\cdots,{\tilde z}_N). 
\end{equation}
Then, due to (\ref{tildez}) and (\ref{jacobian}), we find a duality relation
\begin{equation}
\label{duality-relation}
\rho(z_1,z_2,\cdots,z_k) = \prod_{j=1}^k |{\tilde z}_j|^4 \ {\tilde \rho}({\tilde z}_1,
{\tilde z}_2,\cdots,{\tilde z}_k)
\end{equation}
between the correlation functions $\rho$ defined in (\ref{correlation}) and 
${\tilde \rho}$ defined in (\ref{dual-correlation}). 
\par
For example, when
\begin{equation}
w(z_j) = \left\{ \begin{array}{ll} |z_j|^{\beta \Gamma}, & R \leq |z_j| \leq v, \\ 
0, & {\rm otherwise} \end{array} \right.
\end{equation}
for a real $\Gamma$, we have
\begin{equation}
{\tilde w}({\tilde z}_j) = \left\{ \begin{array}{ll} |{\tilde z}_j|^{- \beta 
\Gamma -\beta (N-1) - 4}, & 
1/v \leq |{\tilde z}_j| \leq 1/R, \\ 0, & {\rm otherwise}. \end{array} \right.
\end{equation}
Therefore, a parameter mapping
\begin{equation}
\label{bridge}
\Gamma \mapsto - \Gamma - N + 1 - \frac{4}{\beta}
\end{equation}
bridges the gap between the correlation functions of the Coulomb gases on the annuli 
$\{ z | \ R \leq |v| \leq v \}$ and $\{ z| \ 1/v \leq |z| \leq 1/R \}$.

\section*{Appendix B} 
\setcounter{equation}{0}
\renewcommand{\theequation}{B.\arabic{equation}}

In this Appendix, we give a proof of the orthogonality relation
\begin{equation}
\label{A-orthogonality}
I_{mn} = \int_0^{2 \pi} |D(z)|^2 p_m({\bar z}) p_n(z)   d \theta = h_n \delta_{m n}, \ \ \ m,n = 0,1,2,\cdots
\end{equation}
($z = r e^{i \theta}$, $r > 1$) with
\begin{equation}
\label{A-dz}
D(z) = \frac{1}{1 - z^{-M}},  \ \ \ M = 1,2,3,\cdots
\end{equation}
for the Type A orthogonal polynomials
\begin{equation}
p_n(z) = \left\{ \begin{array}{ll} z^n, & 0 \leq n < M, \\ 
z^{n - M} ( z^M - 1), & n \geq M \end{array} \right.
\end{equation}
with
\begin{equation}
h_n = \left\{ \begin{array}{ll} \displaystyle 
\frac{2 \pi r^{2 n}}{1 - r^{- 2 M}}, & 0 \leq n < M,  \\ 2 \pi r^{2 n}, & n \geq M \end{array} \right.
\end{equation}
and present a Type B counterpart. The orthogonality relation (\ref{A-orthogonality}) was originally 
proved by Szeg\H{o}\cite{szego}.
\par
\medskip
\noindent
(1) the case $0 \leq m < M$, $0 \leq n < M$
\par
\medskip
\noindent
In this case, due to a series expansion
\begin{equation}
|D(z)|^2 = \frac{1}{\displaystyle 1 - \left( e^{i \theta}/r \right)^M} \frac{1}{\displaystyle 
1 - \left( e^{-i \theta}/r \right)^M}  = \sum_{s=0}^\infty \sum_{t=0}^\infty 
\frac{e^{i M \theta (s - t)}}{r^{M (s + t)}},
\end{equation}
we have
\begin{eqnarray}
I_{mn} & = & \int_0^{2 \pi} |D(z)|^2 {\bar z}^m z^n  d\theta \nonumber \\ 
& = & \sum_{s=0}^\infty \sum_{t = 0}^\infty r^{m + n - M(s + t)}  \int_0^{2 \pi} 
d\theta \  e^{i \theta ( M (s - t) + n - m)}.
\end{eqnarray}
Let us suppose that $M(s - t) + n - m = 0$. Then $-M < M(s-t) = m - n < M$ leading to 
$-1 < s - t < 1$,  which means $s = t$ and $m=n$. Therefore, 
if $s \neq t$ or $m \neq n$, we must have $M(s - t) + n - m \neq 0$. It follows that
\begin{equation}
\int_0^{2 \pi} d\theta \ e^{i \theta ( M (s - t) + n - m)} = 2 \pi \delta_{s t} \delta_{m n}.
\end{equation}
Thus one can see the orthogonality relation
\begin{equation}
I_{mn} = 2 \pi  r^{2 n} \delta_{m n}  \sum_{s=0}^\infty  r^{- 2 M s}
= 2 \pi \frac{r^{2 n}}{\displaystyle 1 - r^{- 2 M}} \delta_{m n}.
\end{equation} 
\par
\medskip
\noindent
(2) the case $m \geq M$, $0 \leq n < M$ 
\par
\medskip
\noindent
It can readily be seen in this case that
\begin{eqnarray}
\label{imn}
I_{m n} & = & \int_0^{2 \pi} |D(z)|^2  {\bar z}^{m - M} ({\bar z}^M - 1) z^n d\theta \nonumber \\ 
& = & r^{m+n} \int_0^{2 \pi} d\theta \frac{1}{1-z^{-M}} e^{i \theta (n-m)} 
\nonumber \\ 
& = & r^{m+n} \sum_{s=0}^\infty r^{- M s} \int_0^{2 \pi} d\theta \ e^{i \theta (n - m - M s)}.
\end{eqnarray}
We suppose that $n - m - Ms = 0$. Then $Ms = n - m <  0$. It follows from $Ms \geq 0$ 
that $n - m - Ms \neq 0$. Therefore the orthogonality $I_{m n} = 0$ is proved. Moreover 
from (\ref{imn}) we obtain the orthogonality relation in the opposite case $0 \leq m < M$, 
$n \geq M$ as
\begin{equation}
\overline{I_{n m}} = \int_0^{2 \pi} |D(z)|^2 {\bar z}^m  z^{n - M} (z^M - 1)  d\theta = 0.
\end{equation}
\par
\medskip
\noindent
(3) the case $m \geq M$, $n \geq M$
\par
\medskip
\noindent
The orthogonality relation in this case is derived as
\begin{eqnarray}
I_{m n} & = & \int_0^{2 \pi} |D(z)|^2 z^{n - M} (z^M - 1)  {\bar z}^{m - M} ({\bar z}^M - 1) d\theta \nonumber \\ 
& = & r^{m + n} \int_0^{2 \pi} d\theta \ e^{i \theta (n - m)} = 2 \pi r^{2 n} \delta_{m n}.
\end{eqnarray} 
\par
In addition, making use of the transformation 
\begin{equation}
z \mapsto \frac{1}{z}
\end{equation}
in the orthogonality relation (\ref{A-orthogonality}), we can derive another 
orthogonality relation for $0 < r <1$ as
\begin{equation}
\label{B-orthogonality}
\int_0^{2 \pi} |D(z)|^2  p_m({\bar z}) p_n(z) d \theta 
= h_n \delta_{m n}, \ \ \  m, n= 0,1,2,\cdots,N-1
\end{equation}
($z = r e^{i \theta}$, $0 < r < 1$) with $D(z)$ in (\ref{A-dz}).
In the case $M < N$, the Type B orthogonal polynomials $p_n(z)$ are given by
\begin{equation}
p_n(z) = \left\{ \begin{array}{ll} z^n (1 - z^M),  & 0 \leq n < N-M, \\ 
z^n, & N-M \leq n <   N \end{array} \right.
\end{equation}
with
\begin{equation}
h_n = \left\{ \begin{array}{ll} 
2 \pi r^{2 n + 2 M}, & 0 \leq n < N - M, \\
\displaystyle 
\frac{2 \pi r^{2 n}}{r^{- 2 M} - 1}, & N - M \leq n < N. 
\end{array} 
\right.
\end{equation}
In the case $M \geq N$, we have
\begin{equation} 
\label{MgeqN}
p_n(z) = z^n, \ \ \ 0 \leq  n < N
\end{equation}
with
\begin{equation}
h_n = 
\displaystyle 
\frac{2 \pi r^{2 n}}{r^{- 2 M} - 1}, \ \ \ 0 \leq n < N.
\end{equation}


\begin{thebibliography}{notitle}

\bibitem{pjf1} P.J. Forrester, {\it Log-Gases and Random Matrices}
(Princeton University Press, 2010).

\bibitem{chafai} D. Chafa\"i, {\it Aspects of Coulomb Gases} (2021) arXiv:2108.10653.

\bibitem{ginibre} J. Ginibre, J. Math. Phys. 6 (1965) 440.

\bibitem{pjf2} P. J. Forrester, Nucl. Phys. B904 (2016) 253.

\bibitem{KS} B.A. Khoruzhenko and H.-J. Sommers, {\it The Oxford Handbook of 
Random Matrix Theory} (ed. by G. Akemann, J. Baik and P. Di Francesco, 
Oxford University Press, 2011) chapter 18.

\bibitem{ZS} K. \.Zyczkowski and H.-J. Sommers, J. Phys. A: Math. Gen. 33 (2000) 2045.

\bibitem{APS} G. Akemann, M.J. Phillips and L. Shifrin, J. Math. Phys. 50 (2009) 063504.

\bibitem{CL} C. Charlier and J. Lenells, Nonlinearity 36 (2023) 1593.

\bibitem{charlier} C. Charlier, Math. Ann. 388 (2024) 3529.

\bibitem{BDM} B. De Bruyne, P.Le Doussal, S.N. Majumdar and G. Schehr, J. Phys. A: 
Math. Theor. 57 (2024) 155002.

\bibitem{BF} S.-S. Byun and P.J. Forrester, {\it Progress on the Study of the Ginibre 
Ensembles} (Springer, 2025).

\bibitem{FBKSZ} J. Fischmann, W. Bruzda, B.A Khoruzhenko, H.-J. Sommers and 
K. \.{Z}yczkowski, J. Phys. A: Math. Theor. 45 (2012) 075203.

\bibitem{AB} G. Akemann and Z. Burda, J. Phys. A: Math. Theor. 45 (2012) 465201.

\bibitem{szego} G. Szeg\H{o}, Trans. Amer. Math. Soc. 37 (1935) 196.

\bibitem{walsh1} J.L. Walsh, Bull. Amer. Math. Soc. 40 (1934) 84.

\bibitem{walsh2} J.L. Walsh, {\it Interpolation and Approximation by Rational Functions 
in the Complex Domain} (American Mathematical Society, 5th edition, 1969).

\bibitem{elliptic} T. Nagao, Phys. Scr. 99 (2024) 125261.

\bibitem{fischmann} J.A. Fischmann, {\it Eigenvalue distributions on a single ring}, 
Ph.D thesis, Queen Mary University of London (2013).

\bibitem{gluck-PRL} M. Gl\"uck, A.R. Kolovsky and H.J. Korsch, 
Phys. Rev. Lett. 82 (1999) 1534.

\bibitem{gluck-PRE} M. Gl\"uck, A.R. Kolovsky and H.J. Korsch, 
Phys. Rev. E60 (1999) 247.

\bibitem{FS} Y.V. Fyodorov and H.-J. Sommers, J. Phys A: Math. Gen. 36 
(2003) 3303.

\bibitem{lubinsky} D.S. Lubinsky, Comput. Methods Funct. Theory {\bf 10} (2010) 135.

\bibitem{CW} J. Cronvall and A. Wennman, {\it A direct approach to soft and hard 
edge universality for random normal matrices} (2025) arXiv:2511.18628.

\bibitem{ACC} Y. Ameur, C. Charlier and J. Cronvall, J. Stat. Phys. 191 (2024) 98.

\bibitem{dyson} F.J. Dyson, J. Math. Phys. 3 (1962) 166.

\bibitem{mehta} M.L. Mehta, {\it Random Matrices} (Elsevier, 3rd edition, 2004).  

\bibitem{jancovici1} B. Jancovici, J. Phys. (Paris) Lett. 42 (1981) L-223.

\bibitem{jancovici2} B. Jancovici, J. Stat. Phys. 28 (1982) 43.

\bibitem{HW} H. Hedenmalm and A. Wennman, Acta. Math. 227 (2021) 309.

\bibitem{DS} A. Dea\~no and N. Simm, Int. Math. Res. Not. 2022 (2022) 210.

\bibitem{ameur} Y. Ameur, {\it A note on normal matrix ensembles at the hard edge} (2018) 
arXiv: 1808.06959.

\bibitem{AKM} Y. Ameur, N.-G. Kang and N. Makarov, Constr. Approx. 50 (2019) 63.

\bibitem{AKMW} Y. Ameur, N.-G. Kang, N. Makarov and A. Wennman, 
J. Funct. Anal. 278 (2020) 108340.

\bibitem{AKS1} Y. Ameur, N.-G. Kang and S.-M. Seo, Anal. Math. Phys. 10 (2020) 68.

\bibitem{seo2020} S.-M. Seo, J. Stat. Phys. 181 (2020) 1473.

\bibitem{seo2022} S.-M. Seo, Ann. Henri Poincar\'e 23 (2022) 2247.

\bibitem{AFLS} M. Allard, P.J. Forrester, S. Lahiry and B. Shen, {\it Partition function 
of 2D Coulomb gases with radially symmetric potentials and a hard wall} (2025) arXiv:2506.14738.

\bibitem{charlier2} C. Charlier, Adv. Math. 408 (2022) 108600.

\bibitem{ACC2} Y. Ameur, C. Charlier and J. Cronvall, Arch. Rational Mech. Anal. 249 (2025) 63.

\bibitem{noda} K. Noda, {\it Partition functions of two-dimensional Coulomb gases with 
circular root- and jump-type singularities} (2025) arXiv:2510.00843.

\bibitem{LY1} S.-Y. Lee and M. Yang, Comm. Math. Phys. 355 (2017) 303.

\bibitem{WW} C. Webb and M.D. Wong, Proc. Lond. Math. Soc. 118 (2019) 1017.

\bibitem{LY2} S.-Y. Lee and M. Yang, Comm. Pure Appl. Math. 76 (2023) 2888.

\bibitem{AKS2} Y. Ameur, N.-G. Kang and S.-M. Seo, Potential Analysis 58 (2023) 331.

\bibitem{byun1} S.-S. Byun, Compt. Methods Funct. Theory 24 (2024) 303.

\bibitem{byun2} S.-S. Byun, E. Yoo, {\it Three topological phases of the elliptic 
Ginibre ensembles with a point charge} (2025) arXiv:2502.02948.

\bibitem{BKSY} S.-S. Byun, N.-G. Kang, S.-M. Seo and M. Yang, J. Lond. Math. Soc. 112 
(2025) e70294.

\bibitem{KKL} M. Kieburg, A.B.J. Kuijlaars and S. Lahiry, Nonlinearity 38 (2025) 065013.

\bibitem{DMMS} A. Dea\~no, K. T-R MacLaughlin, L. Molag and N. Simm, {\it Asymptotics for a 
class of planar orthogonal polynomials and truncated unitary matrices} (2025) arXiv:2505.12633.

\bibitem{BFL} S.-S. Byun, P.J. Forrester and S. Lahiry, {\it Properties of the one-component 
Coulomb gas on a sphere with two macroscopic external charges} (2025) arXiv:2501.05061.

\bibitem{BFKL} S.-S. Byun, P.J. Forrester, A.B.J. Kuijlaars and S. Lahiry, {\it Orthogonal 
polynomials in the spherical ensemble with two insertions} (2025) arXiv:2503:15732.

\bibitem{BM} F. Balogh and D. Merzi, Constr. Approx. 42 (2015) 399.

\bibitem{BGM} F. Balogh, T. Grava and D. Merzi, Constr. Approx. 46 (2017) 109.

\bibitem{BY} S.-S. Byun and M. Yang, SIAM J. Math. Anal. 55 (2023) 6867.

\bibitem{byun3} S.-S. Byun, {\it Anomalous free energy expansions of planar 
Coulomb gases: multi-component and conformal singularity} (2025) arXiv: 2508.00316.

\bibitem{BS} S.-S. Byun and S.-M. Seo, Bernoulli 29 (2023) 1615.

\end{thebibliography}
\end{document}